\documentclass[aps,prl,reprint,preprintnumbers,groupedaddress,nofootinbib]{revtex4-2}
\pdfoutput=1
\usepackage[T1]{fontenc}
\usepackage{lmodern}

\usepackage{pgfplots}
\usepgfplotslibrary{fillbetween}
\usetikzlibrary{calc}
\pgfplotsset{compat=1.18}
\usepackage{booktabs}
\usepackage[dvipsnames,table]{xcolor}
\usepackage{graphicx,amsmath,amssymb,amsthm,multirow,array,bm,esint}
\usepackage{enumitem}
\usepackage[mathscr]{eucal}
\usepackage{amsfonts}
\usepackage{mathtools}
\usepackage{slashed}
\usepackage{hyperref}
\usepackage{upgreek}
\usepackage{comment}
\usepackage{microtype}

\providecommand{\ket}[1]{\left\lvert #1\right\rangle}
\providecommand{\bra}[1]{\left\langle #1\right\rvert}
\providecommand{\braket}[2]{\left\langle #1\middle\vert #2\right\rangle}

\providecommand{\ev}[1]{\left\langle #1\right\rangle}

\hypersetup{
    pdfstartview={FitH},    %
    pdftitle={GPI multi-universe positivity and averaged inner product},
    pdfauthor={Gabriele Di Ubaldo, Luca V. Iliesiu, Henry W. Lin, and Cynthia Yan},
    colorlinks=true,       %
    linkcolor=blue,          %
    citecolor=MidnightBlue, %
    filecolor=magenta,      %
    urlcolor=blue           %
}

\def\mC{\mathcal C}

\newcommand{\es}[2] {\begin{equation} \label{#1} \begin{split} #2 \end{split} \end{equation}}

\newcommand\mR{\mathbb{R}}
\newcommand\mZ{\mathbb{Z}}

\newcommand {\be} {\begin{align}}
\newcommand {\ee} {\end{align}}
\newcommand {\bes} {\begin {equation*}}
\newcommand {\ees} {\end {equation*}}
\newcommand {\beq} {\begin {equation}}
\newcommand {\eeq} {\end {equation}}
\newcommand {\bea} {\begin {eqnarray}}
\newcommand {\ea} {\end {eqnarray}}
\newcommand {\eea} {\end {eqnarray}}

\newcommand{\wormhole}[2]{%
  \pgfmathanglebetweenpoints{\pgfpointanchor{#1}{center}}{\pgfpointanchor{#2}{center}}%
  \let\whangle\pgfmathresult
  \draw[thick, blue!60!black]
    ($(#1.\whangle)+({\whangle+90}:0.35)$)
    .. controls ($(#1)!0.4!(#2)+({\whangle+90}:0.06)$) and ($(#1)!0.6!(#2)+({\whangle+90}:0.06)$) ..
    ($(#2.{\whangle+180})+({\whangle+90}:0.35)$);
  \draw[thick, blue!60!black]
    ($(#1.\whangle)+({\whangle-90}:0.35)$)
    .. controls ($(#1)!0.4!(#2)+({\whangle-90}:0.06)$) and ($(#1)!0.6!(#2)+({\whangle-90}:0.06)$) ..
    ($(#2.{\whangle+180})+({\whangle-90}:0.35)$);
}
\renewcommand{\es}[2]{%
  \begin{equation}%
  \if\relax\detokenize{#1}\relax\else\label{#1}\fi
  \begin{split}#2\end{split}%
  \end{equation}%
}

\def\<{\langle}
\def\>{\rangle}

 \def\ie{\begin{equation}\begin{aligned}}
\def\fe{\end{aligned}\end{equation}}

\usepackage{upgreek}

\def\x{\times}

\def\M{\mathcal{M}}

\def\1{{\rm 1-loop}}

\def\Tr{{\rm Tr}}

\def\c{\cite}

\def\c{\cite}

\def\({\left(}
\def\){\right)}
\def\<{\langle}
\def\>{\rangle}

\def\a{\alpha}

\def\d{\delta}

\def\l{\lambda}

\def\t{\tau}
\def\s{\sigma}
\def\cJ{\mathcal J}

\usepackage{tikz}
\newif\ifsdCompact
\sdCompactfalse
\definecolor{sdink}{HTML}{243440}
\definecolor{sdsurface}{HTML}{EAF0F3}
\definecolor{sdrim}{HTML}{526F80}
\definecolor{sdwick}{HTML}{087F8C}
\definecolor{sdvertex}{HTML}{C24A30}
\definecolor{sdcut}{HTML}{7951A8}
\definecolor{sdmuted}{HTML}{596A76}
\tikzset{
 sd outline/.style={draw=sdink,line width=.85pt,line join=round},
 sd body/.style={sd outline,fill=sdsurface},
 sd rim/.style={draw=sdrim,line width=1pt,fill=white},
 sd wick/.style={draw=sdwick,line width=1.45pt,line cap=round},
 sd interaction/.style={draw=sdvertex,line width=1.45pt,line cap=round},
 sd title/.style={anchor=west,font=\sffamily\bfseries\fontsize{21}{25}\selectfont,text=sdink},
 sd subtitle/.style={anchor=west,font=\sffamily\fontsize{11}{14}\selectfont,text=sdmuted},
 sd formula/.style={font=\fontsize{15}{19}\selectfont,text=sdink},
 sd small/.style={font=\sffamily\fontsize{10}{13}\selectfont,text=sdmuted},
 sd text/.style={font=\sffamily\fontsize{11}{15}\selectfont,text=sdink},
 sd cut/.style={draw=sdcut,line width=1.55pt},
}

\newcommand{\sdMatterLines}[1]{%
 \ifcase#1\relax
 \or
  \draw[sd wick] (0,1.48) .. controls (0,.20) and (.20,0) .. (1.48,0);
  \draw[sd wick] (0,-1.48) .. controls (0,-.20) and (-.20,0) .. (-1.48,0);
 \or
  \draw[sd wick] (0,1.48)--(0,-1.48);
  \draw[draw=sdsurface,line width=5pt] (-.18,0)--(.18,0);
  \draw[sd wick] (-1.48,0)--(1.48,0);
 \or
  \draw[sd wick] (0,1.48) .. controls (0,.20) and (-.20,0) .. (-1.48,0);
  \draw[sd wick] (0,-1.48) .. controls (0,-.20) and (.20,0) .. (1.48,0);
 \or
  \draw[sd interaction] (0,1.48)--(0,-1.48);
  \draw[sd interaction] (-1.48,0)--(1.48,0);
  \fill[sdvertex] (0,0) circle (.105);
 \fi
}
\newcommand{\sdMatterMarks}{%
 \foreach \x/\y in {0/1.48,1.48/0,0/-1.48,-1.48/0}
  \fill[sdink] (\x,\y) circle (.045);
 \node[above,inner sep=2pt] at (0,1.69) {$x_i$};
 \node[above,inner sep=2pt] at (1.70,.40) {$x_j$};
 \node[below,inner sep=2pt] at (0,-1.69) {$x_k$};
 \node[above,inner sep=2pt] at (-1.70,.40) {$x_l$};
}
\newcommand{\sdDisk}[1]{%
 \draw[sd body] (0,0) circle (1.48);
 \sdMatterLines{#1}\sdMatterMarks
}
\newcommand{\sdWormhole}[1]{%
 \path[sd body]
 (-1.60,.40) .. controls (-.65,.40) and (-.40,.65) .. (-.40,1.60)
 --(.40,1.60) .. controls (.40,.65) and (.65,.40) .. (1.60,.40)
 --(1.60,-.40) .. controls (.65,-.40) and (.40,-.65) .. (.40,-1.60)
 --(-.40,-1.60) .. controls (-.40,-.65) and (-.65,-.40) .. (-1.60,-.40)
 --cycle;
 \draw[sd rim] (0,1.60) ellipse (.40 and .12);
 \draw[sd rim] (0,-1.60) ellipse (.40 and .12);
 \draw[sd rim] (-1.60,0) ellipse (.12 and .40);
 \draw[sd rim] (1.60,0) ellipse (.12 and .40);
 \sdMatterLines{#1}\sdMatterMarks
}
\newcommand{\sdEquationSymbols}{%
 \node[sd formula] at (4.85,8.55) {$=$};
 \foreach \x in {9.35,13.85,18.35}
  \node[sd formula] at (\x,8.55) {$+$};
}
\newcommand{\sdMomentTerms}{%
 \node[sd formula] at (2.60,5.85) {$\ell(x_ix_jx_kx_l)$};
 \node[sd formula,text=sdwick] at (7.10,5.85) {$\delta_{ij}\delta_{kl}$};
 \node[sd formula,text=sdwick] at (11.60,5.85) {$\delta_{ik}\delta_{jl}$};
 \node[sd formula,text=sdwick] at (16.10,5.85) {$\delta_{il}\delta_{jk}$};
 \node[sd formula,text=sdvertex] at (20.60,5.85) {$\eta T_{ijkl}$};
 \draw[sdwick,line width=.7pt] (5.55,5.12)--(17.65,5.12);
 \draw[sdwick,line width=.7pt] (5.55,5.12)--(5.55,5.30);
 \draw[sdwick,line width=.7pt] (17.65,5.12)--(17.65,5.30);
 \node[sd small,text=sdwick] at (11.60,4.70) {three Wick pairings};
 \node[sd small,text=sdvertex,align=center] at (20.60,4.75)
  {connected four-point term};
}

\renewcommand{\i}{i}

\newcommand{\bJ}{\mathbf{J}}

\newcommand{\bz}{\mathbf{z}}
\newcommand{\bp}{\mathbf{p}}
\newcommand{\bq}{\mathbf{q}}
\newcommand{\bZ}{\mathbf{Z}}
\newcommand{\bx}{\mathbf{x}}
\newcommand{\brho}{\boldsymbol{\rho}}

\begin{document}

\title{When the gravity path integral describes a statistical average}

\author{Gabriele Di Ubaldo}
\affiliation{Leinweber Institute for Theoretical Physics and Department of Physics, University of California, Berkeley, CA 94720, USA}
\affiliation{RIKEN iTHEMS Center for Interdisciplinary Theoretical and Mathematical Sciences, 2-1 Hirosawa, Wako, Saitama 351-0198, Japan}

\author{Luca V.\ Iliesiu}
\affiliation{Leinweber Institute for Theoretical Physics and Department of Physics, University of California, Berkeley, CA 94720, USA}

\author{Henry W.\ Lin}
\affiliation{Joseph Henry Laboratories and Leinweber Forum for Theoretical Physics,
Princeton University, Princeton, NJ 08544, USA}

\author{Cynthia Yan}
\affiliation{Department of Physics, Harvard University, Cambridge, MA 02138, USA}

\begin{abstract}
We find the necessary and sufficient conditions for the gravitational path integral (GPI) to admit an interpretation in terms of (i) an inner product in the Hilbert space of open/closed universes or (ii) a statistical average of boundary observables. The conditions needed for (i) are necessary but not sufficient for (ii).  
Requiring the stronger condition (ii) places additional constraints on all wormhole amplitudes, giving a concrete diagnostic of whether the GPI computes an average or merely a {\it pseudo-average}.   To exemplify the difference between the two constraints, we present a class of gravitational toy models that satisfy (i) but fail the stronger condition (ii); to emphasize the power of the stronger condition, we initiate a bootstrap study of statistical ensembles based on condition (ii), finding that the resulting constraints substantially sharpen the bounds on the moments of the ensemble obtained by imposing (i).  More broadly, both of these positivity conditions impose nontrivial constraints on wormhole amplitudes and thereby restrict which gravitational effective field theories can be embedded in consistent theories of quantum gravity.

\end{abstract}

\maketitle

\makeatletter
\renewcommand{\l@subsubsection}[2]{}
\makeatother
\tableofcontents
\section{Introduction}

The gravitational path integral (GPI) has become a reliable tool for describing quantum aspects of black hole physics and cosmology~\cite{Gibbons:1976ue}. 
For black holes, it has been used to reproduce the Page curve~\cite{penington2020entanglementwedgereconstructioninformation,Almheiri_2019,Almheiri_2020,penington2020replicawormholesblackhole,Marolf_2021}, to probe the discreteness and fine-grained statistics of the black hole spectrum~\cite{Cotler_2017,saad2019semiclassicalrampsykgravity,Saad:2019lba,stanford2020jtgravityensemblesrandom,saad2019latetimecorrelationfunctions,Cotler_2021,DiUbaldo:2023qli,Maxfield_2021,Boruch:2025ilr,yan2023toruswormholes3dgravity,jafferis2026randommatrixstatistics3d,Saad:2022kfe}  and count black hole microstates~\cite{Dabholkar:2011ec, Dabholkar:2014ema, Iliesiu:2022kny},  or to understand the factorization of the black hole Hilbert space~\cite{harlow2019factorizationproblemjackiwteitelboimgravity,Marolf:2020xie,boruch2024hilbertspacetwosidedblack,Chua_2024,Colafranceschi_2024}. In cosmology, it has been used to study the wavefunction of the universe and to compute transition amplitudes between different spatial geometries~\cite{Hartle:1983ai,Lehners_2023,Chen_2021,ivo2024boundarydensitymatrix,abdalla2026consistentevaluationnoboundaryproposal,cotler2025normnoboundarystate, Usatyuk_2024,usatyuk2025closeduniversesfactorizationensemble,fumagalli2025sitterbraketwormholes,turiaci2025wavefunctionquantums1times,held2025hilbertspacesitterjt}. Underlying all such applications is an assumption about what the GPI computes. Two interpretations of GPI results are widely discussed~\cite{abdalla2026consistentevaluationnoboundaryproposal}:
\begin{enumerate}
    \item The gravitational path integral computes inner products in a Hilbert space of open or closed universes~\cite{Marolf:2020xie,Colafranceschi_2024,witten2025brasketseuclideanpath}. For closed universes, this interpretation is used in cosmology to compute transition amplitudes between different spatial geometries~\cite{Usatyuk_2024,usatyuk2025closeduniversesfactorizationensemble,fumagalli2025sitterbraketwormholes,turiaci2025wavefunctionquantums1times,abdalla2026consistentevaluationnoboundaryproposal}. For open universes, it is used in the context of AdS/CFT to compute inner products between states prepared on different asymptotic slices~\cite{Aharony_2000,witten2007threedimensionalgravityrevisited,Witten_2022,Chandrasekaran:2022eqq,penington2023algebrasstatesjtgravity}.
    \item The gravitational path integral computes an ensemble average, or a suitable coarse-graining, of boundary quantities~\cite{Maldacena:2016hyu,marolf2024natureensemblesgravitationalpath,Marolf:2020xie,deBoer:2023vsm,Engelhardt_2021,belin2025measurespacecftspure,jafferis2025openclosed3dgravityrandom}. For closed universes, it computes an average of boundary observables~\cite{Coleman:1988cy,Giddings:1988cx,Giddings:1988wv,Blommaert_2022,Blommaert:2022ucs,Post_2022}. For open universes, it computes an average of inner products in the boundary theory~\cite{blommaert2020dissectingensemblejtgravity,peng2021babyuniversesensembleaverages,jafferis2023jtgravitymattergeneralized,wang2025wormholesendsworld,hung2025universalstructuresemergentgeometry}. Several of the major recent developments in quantum black hole physics rely on interpreting multi-boundary amplitudes as moments in such an ensemble~\cite{Cotler_2021,Chandra_2022,Belin_2021,Collier_2022,de_Boer_2024,Jafferis_2025,okuyama2020multiboundarycorrelatorsjtgravity,Maloney_2020,Afkhami_Jeddi_2021,Collier_2023,Collier_2024,Aharony_2024,Cotler_2020,Witten_2020,wang2026crossingsymmetryopestatistics,collier2024virasorominimalstring,collier2020universaldynamicsheavyoperators}, including calculations of the Page curve~\cite{penington2020replicawormholesblackhole,Marolf_2021}, statistical counts for the number of black hole states~\cite{balasubramanian2023microscopicoriginentropyblack,boruch2024constructingbpsblackhole}, or statistical arguments for why the Hilbert space of two-sided black holes factorizes~\cite{boruch2024hilbertspacetwosidedblack,Chua_2024,Colafranceschi_2024}. While for open universes this approach has thus led to a detailed understanding of the quantum properties of black holes, 
    for closed universes, the statistical interpretation has led to perplexing results: this interpretation implies that the closed-universe Hilbert space in a fixed $\alpha$-sector is one-dimensional~\cite{Marolf:2020xie,mcnamara2020babyuniversesholographyswampland,iliesiu2024nonperturbativebulkhilbertspace,Usatyuk_2024,usatyuk2025closeduniversesfactorizationensemble,liu2025filteringcftslargen}, motivating several proposals for how to describe non-trivial physics seen by an observer inside the closed universe~\cite{harlow2025quantummechanicsobserversgravity,abdalla2025gravitationalpathintegralobservers,Chandrasekaran_2023,witten2023algebrasregionsobservers,witten2023backgroundindependentalgebraquantum,blommaert2025absoluteentropyobserversnoboundary,Jensen_2023}.
\end{enumerate}
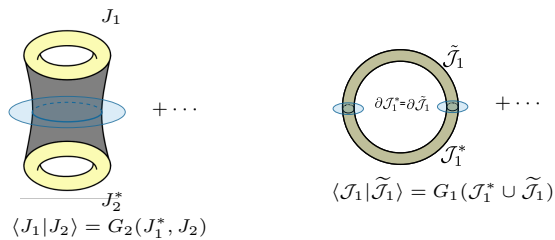
\begin{figure}[t!]
\centering
\begin{minipage}[c]{0.49\columnwidth}
  \centering
  \begin{tikzpicture}[baseline=(wh.base)]
    \node[inner sep=0] (wh)
      {\includegraphics[width=0.38\linewidth]{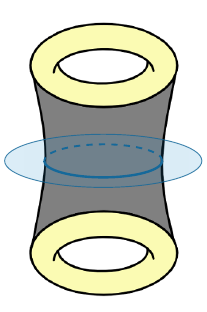}};
    \node[anchor=west,font=\scriptsize] at ([xshift=-1.5em]wh.north east) {$J_1$};
    \node[anchor=west,font=\scriptsize] at ([xshift=-1.5em]wh.south east) {$J_2^*$};
    \node[anchor=west,font=\scriptsize] at ([xshift=0.6em]wh.east) {$+\cdots$};
  \end{tikzpicture}\\[-0.15em]
  {\scriptsize $\langle J_1\vert J_2\rangle=G_2(J_1^*,J_2)$}
\end{minipage}
\hfill
\begin{minipage}[c]{0.49\columnwidth}
  \centering
  \begin{tikzpicture}[baseline=(open.base)]
    \node[inner sep=0] (open)
      {\includegraphics[width=0.44\linewidth]{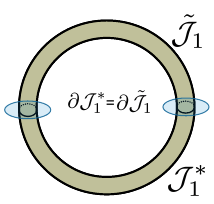}};
    \node[anchor=west,font=\scriptsize] at ([xshift=0.6em]open.east) {$+\cdots$};
  \end{tikzpicture}\\[-0.05em]
  {\scriptsize $\langle\cJ_1\vert\widetilde{\cJ}_1\rangle
  =G_1(\cJ_1^*\cup\widetilde{\cJ}_1)$}
\end{minipage}
\caption{Examples of geometries contributing to closed- and open-universe overlaps. The blue cuts indicate a closed slice (left) and an open slice (right).}
\label{fig:closed-open-overlaps}
\end{figure}
Given the importance of these developments, understanding whether interpretation (1), interpretation (2), both, or neither is valid is critical for making further progress. In this paper, we present the necessary and sufficient positivity criteria for each interpretation.\footnote{Note that even in factorizing theories of gravity these criteria will be satisfied, albeit trivially.}

For interpretation (1), the relevant rules follow from requiring that every linear combination of states prepared by the path integral has non-negative norm~\cite{Glimm:1987ylb,Colafranceschi_2024,witten2025brasketseuclideanpath,neeb2018reflectionpositivityarepresentationtheoretic}. This requirement translates into the positive semi-definiteness of all Gram matrices of closed- or open-universe states. The resulting inequalities place severe constraints on single- and multi-boundary amplitudes computed by the GPI, not manifestly satisfied by an arbitrary low-energy theory of gravity~\cite{Colafranceschi:2023txs,Maloney_2010,Keller_2015,benjamin2020puregravityconicaldefects,loges202310dconstructioneuclideanaxion,chen2026negativeshocksversusstatic}.

The conditions required for interpretation (1) are also necessary for interpretation (2), but they are not sufficient. The distinction can be formulated as a moment problem~\cite{schmüdgen2020lecturesmomentproblem,lasserre2018momentsoshierarchy}. For any finite collection of boundary conditions, the GPI amplitudes define a linear functional on polynomials in the corresponding boundary observables: each monomial can be replaced by the wormhole amplitude with the same collection of boundary insertions. If these amplitudes arise from a statistical ensemble, the resulting functional is the ensemble average of the polynomial and must therefore be non-negative for polynomials that are non-negative at every point.

Hilbert-space positivity tests a smaller class of polynomials. A polynomial is called a sum of squares (SOS) if it can be written as a sum of absolute squares of other polynomials~\cite{Blekherman_2021,Huber_2024}. This representation makes its pointwise non-negativity manifest, and its GPI average can be written as a sum of Hilbert-space norms; consequently, the positivity of the GPI linear functional applied to such polynomials is guaranteed by the positivity of all norms in the closed universe Hilbert space. However, for more than two variables, not every pointwise non-negative polynomial is SOS~\cite{Blekherman_2021}. Positive polynomials that are not SOS therefore give additional constraints on wormhole amplitudes which cannot be obtained from the positivity of any Gram matrix. The simplest examples of such polynomials yield concrete constraints on four-boundary wormhole amplitudes, not implied by the positivity of any norm in the closed universe Hilbert space.

In the computer science literature, a normalized linear functional that is positive on every SOS polynomial is called a {\it pseudo-expectation}, and its formal moment data are sometimes described as a {\it pseudo-distribution}~\cite{barak2014sumofsquaresproofsquestoptimal,keren2026stationarityenoughtightnessquantum}. A pseudo-expectation obeys the positivity conditions visible to Hilbert-space norms but may not admit a representation as an integral against any positive measure. Thus, the moments satisfying interpretation (1) define a pseudo-expectation, while interpretation (2) requires %
a genuine probability distribution. Thus, if the GPI only satisfies (1) and not (2), we say that it computes a pseudo-average.

Open-universe positivity imposes additional constraints. In the statistical interpretation, the random variables associated with boundary partition functions cannot take arbitrary values: in each member of the ensemble, they must themselves define positive open-universe inner products. Their support is consequently restricted to a non-trivial domain $K$~\cite{schmüdgen2020lecturesmomentproblem,Putinar_1999,lasserre2018momentsoshierarchy}.{ For example, the random variable that represents a thermal partition function is restricted to positive real numbers for positive temperatures.} The GPI must then be positive on every polynomial that is non-negative on $K$, including polynomials of low degree that need not be positive away from $K$. Thus, in this case, the distinction between SOS and non-SOS polynomials is not by itself sufficient to distinguish a Hilbert-space interpretation from a statistical one. Nevertheless, we construct concrete examples of constraints that the statistical interpretation needs to satisfy but the conventional interpretation does not, solely involving two-boundary wormhole amplitudes. Such amplitudes are typically easier to compute than genuine multi-boundary wormhole amplitudes, since two-boundary saddles generally enjoy additional isometries. 
It can therefore be easier to disprove open-universe statistical positivity than closed-universe statistical positivity, whose distinguishing constraints involve wormholes with a larger number of boundaries.

The remainder of this paper is organized as follows. In section~\ref{sec:constraints-from-closed-universes}, we derive the Hilbert-space and statistical positivity conditions for closed universes and explain their relation to SOS and non-SOS polynomials. We also provide a gravitational toy model that has a well-defined closed-universe Hilbert space interpretation but violates the statistical positivity conditions and therefore does not have a statistical interpretation. In section~\ref{sec:constraints-from-open-universes}, we repeat this analysis for open universes, where the support of the putative statistical measure is restricted by Hilbert space positivity in each boundary theory. In section~\ref{sec:positivity-cones}, we summarize the relations between the resulting positivity cones. We conclude in section~\ref{sec:discussion} with a discussion about the relation to the Marolf-Maxfield construction of the closed-universe Hilbert space, about possible applications to gravitational effective theories, and about how statistical positivity can be used when bootstrapping statistical models. In appendix~\ref{sec:pseudo-average-examples}, we construct explicit closed- and open-universe pseudo-averages that obey Hilbert-space positivity while violating statistical positivity, and discuss gravitational realizations of such constructions. In appendix~\ref{sec:what-fails-in-the-MM-construction}, we use a quantum mechanical example to explain how commuting operators can fail to admit a joint spectral representation and consequently need not obey non-SOS positivity. Finally, in appendix~\ref{sec:non-SOS-bootstrap}, we compare Hilbert-space and statistical positivity in bootstrap studies of multivariate probability distributions and multi-matrix integrals, finding that the latter significantly sharpens the resulting bounds.

\section{Constraints from closed universes}
\label{sec:constraints-from-closed-universes}

We begin by fixing notation. Let
\es{eq:def-Gn}{
G_n(J_1,\ldots,J_n)
    = \int_{\partial M=\bigsqcup_{i=1}^n B_i}
      \mathcal D g\,\mathcal D\Phi\,e^{-I[g,\Phi]}
}
denote the gravitational path integral over spacetimes whose asymptotic boundary has $n$ connected components. The boundary conditions for the metric and for all matter fields on $B_i$ are collectively denoted by $J_i$. The integral in \eqref{eq:def-Gn} includes the prescribed sum over bulk topologies~\cite{belin2026universalsumtopologies3d}, and we will often suppress the subscript $n$. We denote by $J^*$ the reflected, or CPT-conjugate, boundary condition. For a collection $\bJ=(J_1,\ldots,J_m)$, we write $\bJ^*=(J_1^*,\ldots,J_m^*)$; because the asymptotic boundaries are unlabeled, their ordering in $G$ will be immaterial.

\subsection{When the GPI computes an inner product in the closed-universe Hilbert space}
\label{sec:closed-universe-conventional-interpretation}

In the conventional Hilbert-space interpretation, a collection of closed boundaries $\bJ=(J_1,\ldots,J_m)$ prepares a state $\ket{\bJ}$ in the closed-universe Hilbert space. The state may itself contain an arbitrary number of disconnected components. Its overlap with a second state is
\es{eq:closed-inner-product}{
G(\bJ^*,\bJ')=\braket{\bJ}{\bJ'}\,.
}
This prescription defines a pre-Hilbert space if and only if, for every finite collection of states $\ket{\bJ_a}$ and coefficients $c_a$,
\es{eq:closed-gram-positivity}{
\sum_{a,b}c_a^*c_b\,G(\bJ_a^*,\bJ_b)
=\left\lVert\sum_a c_a\ket{\bJ_a}\right\rVert^2\geq0\,.
}
Equivalently, every matrix $G(\bJ_a^*,\bJ_b)$ must be positive semi-definite. To then define a Hilbert space, null states can be quotiented out to obtain Gram matrices that are positive definite.

The simplest constraint follows from imposing that the Gram matrix of two one-component states is PSD. Its determinant gives
\es{eq:example-2bdy-wormhole-inequality}{
G_2^\text{conn}(J_1^*,J_1)G_2^\text{conn}(J_2^*,J_2)
    \geq \left|G_2^\text{conn}(J_1^*,J_2)\right|^2\,.
}
where, for convenience, we have added a subscript to indicate the total number of boundary-connected components that the GPI is evaluated on and a superscript to indicate that only connected wormhole amplitudes contribute.  Eq.~\eqref{eq:example-2bdy-wormhole-inequality} thus constrains two boundary wormhole amplitudes. 
The same condition applied to states with two disconnected components gives, for example,
\es{eq:example-4bdy-wormhole-inequality}{
&G_4(J_1^*,J_2^*,J_1,J_2)
 G_4(J_3^*,J_4^*,J_3,J_4)\\
&\hspace{2.8cm}\geq
\left|G_4(J_1^*,J_2^*,J_3,J_4)\right|^2\,,
}
which involves amplitudes of not only 2, but also 3 and 4-boundary wormholes. 
Higher-rank Gram matrices give an infinite family of analogous constraints involving amplitudes with arbitrarily many boundaries.

\subsection{When the GPI computes a statistical average over boundary observables}
\label{sec:closed-universe-statistical-interpretation}

Can the same collection of amplitudes be interpreted as moments of boundary observables in a distribution with positive measure? This requires
\es{eq:statistical-interpretation}{
&G(J_1^*,\ldots,J_k^*,J'_1,\ldots,J'_{k'})
 =\braket{J_1,\ldots,J_k}{J'_1,\ldots,J'_{k'}}\\
&\qquad =\int d\mu(\alpha)\,
 Z_\alpha[J_1]^*\cdots Z_\alpha[J_k]^*
 Z_\alpha[J'_1]\cdots Z_\alpha[J'_{k'}],
}
where $d\mu(\alpha)$ is a positive measure and $Z_\alpha[J^*]=Z_\alpha[J]^*$. The label $\alpha$ can denote an ensemble member, an $\alpha$-state, or more generally the microscopic data retained by the coarse-grained description.

For a fixed collection of boundary conditions $J_1,\ldots,J_k$, consider a real-valued polynomial on $\mathbb C^k$,
\es{eq:hermitian-polynomial}{
P(\bz,\bar\bz)
=\sum_{\bp,\bq}c_{\bp,\bq}^P\,
  \bz^{\bp}\bar\bz^{\bq},
\qquad
(c_{\bp,\bq}^P)=(c_{\bq,\bp}^P)^*\,.
}
Here $\bp,\bq\in\mathbb Z_{\geq0}^k$ and $\bz^{\bp}=\prod_i z_i^{p_i}$. We associate $z_i$ with $J_i$ and $\bar z_i$ with $J_i^*$. The GPI then defines a linear functional on such polynomials,
\es{eq:def-LG}{
L_G[P]
=\sum_{\bp,\bq}c_{\bp,\bq}^P\,
G\!\left(\mathbf J_{\bp,\bq}\right),
}
where
\es{eq:def-boundary-multi-index}{
\mathbf J_{\bp,\bq}
=\left(
\underbrace{J_1,\ldots}_{p_1\,\text{copies}},\ldots,
\underbrace{J_k,\ldots}_{p_k\,\text{copies}},
\underbrace{J_1^*,\ldots}_{q_1\,\text{copies}},\ldots,
\underbrace{J_k^*,\ldots}_{q_k\,\text{copies}}
\right).
}
If \eqref{eq:statistical-interpretation} holds, then
\es{eq:LG-as-average}{
L_G[P]=\int d\mu(\alpha)\,
P\bigl(\bZ_\alpha,\bar\bZ_\alpha\bigr).
}
Consequently,
\es{eq:inequality-satisfied-in-stat-interpretation}{
L_G[P]\geq0
\qquad\text{for every }P\geq0\text{ on }\mathbb C^k\,,
}
where, by convention, we normalize the path integral such that $L_G[1]=1$. 
Conversely, after regarding the real and imaginary parts of $z_i$ as $2k$ real variables, the Riesz--Haviland theorem implies that \eqref{eq:inequality-satisfied-in-stat-interpretation}, imposed for all polynomials, is sufficient for $L_G$ to admit a representing positive measure~\cite{schmüdgen2020lecturesmomentproblem,Putinar_1999}. Thus, this is the necessary and sufficient criterion for the closed-universe statistical interpretation.

The relation to Hilbert-space positivity is immediate for a polynomial $P$ that can be written as a sum of squares of other polynomials $Q_a$,
\es{eq:definition-SOS-polynomial}{
P(\bz,\bar\bz)=\sum_a\left|Q_a(\bz)\right|^2\,.
}
Such polynomials are known as SOS polynomials. Indeed,
\es{eq:SOS-is-norm}{
L_G[P]
=\sum_a\left\lVert \sum_{\bp,\bq} c_{\bp,\bq}^{Q_a} \ket{\bJ_{\bp,\bq}}\right\rVert^2
\geq0\,.
}
Requiring $L_G[P]\geq0$ for all Hermitian SOS polynomials is therefore equivalent to the Gram-matrix condition \eqref{eq:closed-gram-positivity}. We denote the corresponding cone of test polynomials by
\es{eq:closed-Hilbert-cone}{
\mC_{\rm closed}^{\rm Hilbert}
=\bigcup_{k\geq1}\Sigma(\mathbb C^k),
}
where $\Sigma(\mathbb C^k)$ is the cone of polynomials of the form \eqref{eq:definition-SOS-polynomial}. It is a convex cone because it is closed under addition and multiplication by non-negative real numbers.

Not every pointwise non-negative polynomial is SOS~\cite{Blekherman_2021,Huber_2024}. For non-homogeneous real polynomials, the first examples occur at degree six in two variables and at degree four in three variables; equivalently, one can use homogeneous ternary sextics or quaternary quartics. Therefore, the cone
\es{eq:closed-stat-cone}{
\mC_{\rm closed}^{\rm stat}
=\bigcup_{k\geq1}\operatorname{Pos}(\mathbb C^k)
}
of all pointwise non-negative polynomials is strictly larger than \eqref{eq:closed-Hilbert-cone}. Positivity on this larger cone is not implied by the positivity of closed-universe norms.

For a concrete example, take two CPT-invariant boundary conditions $J_1,J_2$, so that the associated variables $x_1,x_2$ are real.\footnote{If such boundary conditions do not exist one can, for instance, instead identify  $x_i\leftrightarrow Z_\a[J_i]+ Z_\a[J_i^*]$.} The Motzkin polynomial
\es{eq:motzkin-polynomial}{
P_M(x_1,x_2)
=x_1^4x_2^2+x_1^2x_2^4+1-3x_1^2x_2^2
\geq0
}
is famously non-SOS~\cite{ChoiLamReznick}. Statistical positivity consequently requires
\es{eq:motzkin-GPI-inequality}{
&G_6(J_1,J_1,J_1,J_1,J_2,J_2)
+G_6(J_1,J_1,J_2,J_2,J_2,J_2)+G_0\\
&\hspace{2.5cm}-3G_4(J_1,J_1,J_2,J_2)\geq0\,.
}
Here $G_0=L_G[1] = 1$, which can be set to one when the statistical measure is normalized. To constrain wormhole amplitudes with a smaller number of boundaries, we need to increase the number of possible boundary conditions. For four CPT-invariant boundary conditions, the Choi--Lam polynomial~\cite{ChoiLamReznick}
\es{eq:choi-lam-polynomial}{
P_{CL}(x_1,x_2,x_3,x_4)
&=x_1^2x_2^2+x_2^2x_3^2+x_3^2x_1^2+x_4^4\\ 
&-4x_1x_2x_3x_4
\geq0
}
implies
\es{eq:choi-lam-GPI-inequality}{
&G_4(J_1,J_1,J_2,J_2)
+G_4(J_2,J_2,J_3,J_3)
+G_4(J_3,J_3,J_1,J_1)\\
&\qquad+G_4(J_4,J_4,J_4,J_4)
-4G_4(J_1,J_2,J_3,J_4)\geq0\,.
}
Neither \eqref{eq:motzkin-GPI-inequality} nor \eqref{eq:choi-lam-GPI-inequality} follows from the positive semi-definiteness of the closed-universe Gram matrix alone. More abstractly, at any fixed degree for which the non-negative and SOS cones differ, a separating linear functional gives truncated GPI data that are non-negative on every SOS polynomial but negative on a pointwise non-negative polynomial.

There are several important cases in which the conditions in this section automatically follow from those in section~\ref{sec:closed-universe-conventional-interpretation}. The first is the disk and cylinder approximation, in which all connected wormhole amplitudes with more than two boundaries vanish. If the two-boundary Gram condition \eqref{eq:example-2bdy-wormhole-inequality} holds, this covariance is positive semi-definite and defines a positive Gaussian measure. All polynomial positivity constraints, including the non-SOS constraints above, are then automatically satisfied. If higher-boundary connected amplitudes are merely suppressed rather than zero, this conclusion is perturbative: sufficiently high-degree polynomials, whose degree scales with $G_N^{-1}$, or low-degree polynomials whose variables are associated with boundary conditions that also scale with $G_N^{-1}$ can still be sensitive to the suppressed non-Gaussian cumulants.

A second sufficient condition follows by restricting the growth of the moments. If $L_G$ is normalized, positive on all SOS polynomials, and satisfies  Carleman's condition~\cite{putinar2008multivariatedeterminateness,schmüdgen2020lecturesmomentproblem},
\es{eq:multivariate-Carleman}{
\sum_{n=1}^{\infty}
\left(L_G[x_a^{2n}]\right)^{-1/(2n)}=\infty,
\qquad a=1,\ldots,d.
}
for every identification $x_a \leftrightarrow Z_\alpha[J_a]$, then there exists a unique positive measure that reproduces all the moments $L_G[x_a^{2n}]$. Thus, in this case, Hilbert-space positivity is sufficient for statistical positivity and all non-SOS constraints are automatically satisfied.

The Carleman condition is only sufficient, and it can fail when the high moments grow too rapidly. This caveat is important in gravity where we can in principle assess whether \eqref{eq:multivariate-Carleman} is satisfied. This was studied in~\cite{Janssen_2021} for JT gravity. For moments associated to $Z_\alpha[\beta]$ where $\beta \sim O(1)$, high moments~\cite{hide2025largenasymptoticsweilpeterssonvolumes} receive the largest contribution from surfaces with arbitrarily large genus $g$;\footnote{This is because $\text{Vol}_{g,n}(\mathbf b) \sim (n!)n^g$~\cite{manin1999invertiblecohomologicalfieldtheories}.} since the sum over genus is asymptotic understanding whether the Carleman condition is satisfied requires a resummation of the entire expansion which is unknown. However, in the Airy regime where $\beta\sim e^\frac{2S_0}3$, the resummation over genus can be exactly performed and the moments can thus be explicitly computed. In this regime, the $n$-boundary moments grow as $\exp[c n^3+o(n^3)]$ for some $c>0$. Consequently, the terms in \eqref{eq:multivariate-Carleman} decay as $\exp[-4c n^2+o(n^2)]$ and the Carleman sum converges. This does not imply that JT gravity has no statistical interpretation, but it means that such an interpretation cannot be inferred from SOS positivity together with the Carleman criterion. More generally, in the absence of an independent construction of the underlying ensemble, gravitational theories with such rapidly growing multi-boundary amplitudes require an explicit check of the non-SOS positivity constraints.

\subsection{A gravitational model without a statistical interpretation}\label{sec:toymodel}

To make things explicit, it is useful to provide an explicit example of wormhole amplitudes, $G_n(J_1, \dots, J_n)$, consistent with section~\ref{sec:closed-universe-conventional-interpretation} but inconsistent with a statistical interpretation. We construct such an example in appendix \ref{sec:closed-universe-Hilbert-space-positivity--without-statistical-positivity}. Although this example assigns concrete values to all amplitudes, their gravitational origin is unclear. We therefore seek a toy gravitational theory whose GPI has the same positivity properties.

Our strategy is to couple a gravitational theory that admits an ensemble interpretation to interacting matter whose correlators satisfy reflection positivity but need not satisfy statistical positivity. We then ask whether the violation of statistical positivity in the matter theory persists after coupling to gravity, preventing the coupled theory from admitting an ensemble interpretation.

We implement this strategy in a toy model described in detail in Appendix \ref{sec:a-grav-example}. We start with the Marolf-Maxfield topological model~\cite{Marolf:2020xie}, a $2d$ model of quantum gravity that assigns to surfaces of genus $g$ with $n$ boundaries the action, 
\es{eq:Marolf-Maxfield-action}{
I[g_{\mu\nu}] = -S_0(2-2g)\,.
}
To compute the GPI with  $n$ boundaries (where we label each boundary by $Z$), we sum over the possible partitions $\pi \in \Pi_n$ of the boundaries, with each block corresponding to a connected component of spacetime, 
\es{eq:MM-GPI-Gn}{
L_G[Z^n] &= G_n(Z, \dots, Z) = \sum_{\pi \in \Pi_n} \lambda^{|\pi|}\,, \\
\l&=\sum_{g\ge0}e^{S_0(2-2g)}
 =\frac{e^{2S_0}}{1-e^{-2S_0}},
} 
where $|\pi|$ is the number of connected components carrying the specified boundaries, and $\lambda$ captures the sum over genera for each component. 
By itself, this model has an ensemble interpretation, whose moments \eqref{eq:MM-GPI-Gn} are computed by
\es{eq:MM-GPI-ensemble}{
L_G[Z^n] = \sum_{\alpha=0}^\infty p_\alpha (Z_\alpha)^n, \,\,\,\,\,p_\alpha \equiv \frac{e^{-\lambda}\lambda^\alpha}{\alpha!}, \,\,\,\, \, Z_\alpha=\alpha.
}
We now couple this theory to a simplified model of interacting matter fields $\widehat O_1$, \dots, $\widehat O_r$. On each connected component, the matter correlators are independent of the genus, the number of boundaries, the assignment of insertions to boundaries, and their ordering. Additionally, correlators factorize between disconnected components. We encode these correlators with a linear functional $\ell$ acting on polynomials in commuting\footnote{From the boundary point of view, these matter correlators satisfy $[\widehat O_r, \widehat O _{s}] = 0$. This is simpler than the most general possibility for matter operators in a topological quantum mechanics, see e.g. \cite{Lin:2022rzw}.} variables $(x_1,\ldots,x_r)$, with each monomial specifying the corresponding matter insertions $(\widehat O_1,\ldots, \widehat O_r)$. Specifically, we take $\ev{ \widehat{O}_{1}^{i_1} \cdots  \widehat{O}_{r}^{i_r}} = \ell(x_1^{i_1} \cdots x_{r}^{i_r})$.\footnote{For a single boundary with some number of insertions $n$, $\ell(x_1 \cdots x_n) \propto \Tr \, (\widehat O_1 \cdots \widehat O_n)$. More generally, $\ell$ is not necessarily proportional to a trace in the boundary Hilbert space; see Figure \ref{fig:wormhole4pt}.} {These rules are illustrated in Figures \ref{fig:disk4point} and \ref{fig:wormhole4pt}, where we list the 4-point function of such operators on the disk and on a 4-boundary wormhole, respectively.  }

We choose $\ell$ to be normalized and to satisfy reflection positivity:
\begin{equation}
 \ell(1)=1,\qquad \ell(P^*P)>0\quad(P\ne0),
 \label{eq:positivity-for-matter}
\end{equation} 
for any polynomial $P(x_1, \dots, x_r)$. This is the same SOS positivity criterion discussed for the functional $L_G$ above.  We can, however, choose $\ell$ to violate non-SOS positivity; for instance, we can choose $\ell(p_{CL}(x))<0$ for the Choi-Lam polynomial defined in \eqref{eq:choi-lam-polynomial}.

What happens when we couple such matter to the Marolf-Maxfield model? Each connected genus-$g$ surface with $n$ boundaries on which we insert a set of operators associated to the polynomials $P_1,\ldots,P_n$ is now assigned the weight
\begin{equation}
 G_{g,n}^\text{conn}(P_1,\ldots,P_n)
 =e^{S_0(2-2g)}\,\ell(P_1\cdots P_n).
 \label{eq:app-connected}
\end{equation}
Summing over genera and partitions of the $n$ boundaries as in \eqref{eq:MM-GPI-Gn}, we obtain
\begin{equation}
  G_n(P_1,\ldots,P_n)
 =\sum_{\pi\in\Pi_n}\lambda^{|\pi|}
   \prod_{B\in\pi}\ell\!\left(\prod_{a\in B}P_a\right).
 \label{eq:app-partitions}
\end{equation}
In Appendix \ref{sec:a-grav-example} we show that the reflection positivity of the matter correlators \eqref{eq:positivity-for-matter} implies that the $n$-boundary amplitudes satisfy all SOS-positivity conditions; therefore, the GPI indeed computes inner products in a well-defined Hilbert space.

Similarly, the appendix also shows that a violation of statistical positivity in the matter sector persists after coupling to gravity. To get some intuition, we first choose an explicit example for the linear functional $\ell$, for which the matter four-point correlators are determined by 
 \es{}{\ell(x_ix_jx_kx_l)&=\delta_{ij}\delta_{kl}+\delta_{ik}\delta_{jl}+\delta_{il}\delta_{jk}+ \eta T_{ijkl},
 }
 where the tensor $\eta \,T$ encodes the departure from Wick contractions and is chosen so that $\ell(p_{CL}(x))<0$. When we appropriately scale the elements by setting $\eta  \sim O(e^{2S_0})$, we show that $L_G[p_{CL}(Z(x_1), \dots, Z(x_4))] < 0$, where each $Z(x_i)$ corresponds to one boundary on which we insert the operator $O_i$. This guarantees that the gravitational theory does not have a statistical interpretation. 
 
 In a more realistic theory of gravity, particle interactions are, of course, not expected to scale as $O(e^{2S_0}) \sim O(e^{L^{d-2}/G_N})$, where $L$ is the characteristic length scale of a GPI saddle. In such theories, $L_G[p_{CL}(Z(x_1), \dots, Z(x_4))]$ might therefore be positive; nevertheless, as long as there exists a non-SOS positive polynomial $p_\text{non-SOS}(x)$ for which $\ell(p_\text{non-SOS}(x))<0$, we show that there exists a real one-variable positive polynomial $s(z)$ such that 
 \es{}{
 L_G\left[s(Z(1)) p_\text{non-SOS}(Z(x))\right]< 0\,,
 }
 where each $Z(1)$ corresponds to one boundary with no matter insertion. Since $W(z, x) = s(z) p_\text{non-SOS}(x)$  is itself a non-SOS positive polynomial, statistical positivity is violated. Thus, at least for the Marolf-Maxfield model, any coupling to an interacting matter theory whose correlators violate non-SOS positivity ruins its statistical interpretation. It would be interesting to determine whether the same conclusion holds in other gravitational theories with known ensemble interpretations, such as JT gravity.

\section{Constraints from open universes}
\label{sec:constraints-from-open-universes}

\subsection{When the GPI computes an inner product in the open-universe Hilbert space}
\label{sec:open-universe-conventional-interpretation}

An open-universe state is prepared by a path integral whose boundary includes an open slice $\cJ$. We denote the resulting state by $\ket{\cJ}$. Let $\cJ^*$ be the reflected, or CPT-conjugate, slice. Whenever the codimension two boundaries agree, $\partial\cJ^*=\partial\cJ'$, the two slices can be glued to form a closed asymptotic boundary~\cite{hartman2025triangulatingquantumgravityads3,hartman2025conformalturaevvirotheory,jafferis2026facetshyperbolictetrahedronopen}
\es{eq:open-glued-boundary}{
J=\cJ^*\cup_{\partial\cJ}\cJ'\,.
}
The GPI then computes
\es{eq:open-inner-product}{
G(J)=G(\cJ^*\cup\cJ')=\braket{\cJ}{\cJ'}\,.
}
Thus, for every finite set of mutually compatible open slices, the kernel
\es{eq:open-gram-positivity}{
G(\cJ_a^*\cup\cJ_b)\succeq0\,,
}
must be positive semi-definite.\footnote{Note that if $\partial \cJ_a^* = \partial  \cJ_b = \emptyset$ the inner products become the closed universe inner products and \eqref{eq:open-gram-positivity} becomes \eqref{eq:definition-SOS-polynomial}. We thus define the open-universe constraint as those obtained by taking  $\partial \cJ_a^* = \partial  \cJ_b \neq \emptyset$.  } As for closed universes, this statement means that the norm of every linear combination of open-universe states is non-negative. The additional power of \eqref{eq:open-gram-positivity} comes from the fact that the same set of codimension-two boundaries can often be connected by the open slices in inequivalent ways.

To illustrate these conditions, it is useful to discuss two examples. First, consider asymptotically AdS states prepared by a boundary with topology $I\times S^{d-1}$~\cite{giombi2008onelooppartitionfunctions3d,hartman2014universalspectrum2dconformal}. We can fix the radius of $S^{d-1}$ to one and let $\tau$ be the length of the interval $I$. Gluing states associated to intervals of lengths $\tau$ and $\tau'$ gives a boundary with topology $S^1\times S^{d-1}$, whose circle has a length $\tau+\tau'$. Therefore,
\es{eq:single-open-Hankel-kernel}{
G_1(\tau+\tau')=\braket{\tau}{\tau'}\succeq0\,.
}
Under standard regularity assumptions, positivity of this Hankel kernel gives a bilateral Laplace transform of a positive measure. If we additionally demand that the generator of Euclidean evolution be bounded below and shift the lower edge of its spectrum to $E=0$, the measure must be supported on $\mathbb R_+$~\cite{johnson2020nonperturbativejtgravity,johnson2022consistencyconditionsnonperturbativecompletions}. The resulting condition is that $G_1$ is completely monotonic~\cite{Scott_2014,kozhasov2019positivitycertificatesintegralrepresentations},
\es{eq:complete-monotonicity}{
(-1)^m\partial_\beta^mG_1(\beta)\geq0 
\text{ for all }m\in \mZ_+ \text{ and } \beta \in \mR,
}
or, equivalently, to a Laplace representation
\es{eq:positive-Laplace-representation}{
G_1(\beta)=\int_0^\infty dE\,\rho_1(E)e^{-\beta E}, \text{ with }
\qquad \rho_1(E)\geq0\,.
}
Without the lower-boundedness assumption, one still needs $\rho_1(E) \geq 0$ without requiring \eqref{eq:complete-monotonicity}.

One can similarly consider states with $n$ disconnected components of topology $I\times S^{d-1}$. If the intervals in the bra and ket connect the codimension-two boundaries in the same way, positivity requires
\es{eq:G-multi-open}{
G_n(\tau_1+\tau'_1,\ldots,\tau_n+\tau'_n)
=\braket{\tau_1,\ldots,\tau_n}
         {\tau'_1,\ldots,\tau'_n}\succeq0\,.
}
After shifting each spectral edge to zero, this is equivalent to joint complete monotonicity~\cite{Scott_2014},
\es{eq:multivariate-complete-monotonicity}{
(-1)^{|\boldsymbol\ell|}
\partial_{\beta_1}^{\ell_1}\cdots
\partial_{\beta_n}^{\ell_n}
G_n(\beta_1,\ldots,\beta_n)\geq0
}
for every multi-index $\boldsymbol\ell\in\mathbb Z_{\geq0}^n$ and all $\beta_i \in \mathbb R$. Equivalently,
\es{eq:multi-density-representation}{
G_n(\boldsymbol\beta)
&=\int_{\mathbb R_+^n}\prod_{i=1}^n dE_i\,
\rho_n(E_1,\ldots,E_n)
e^{-\sum_i\beta_iE_i},\\
&\text{ where }  \rho_n(E_1,\ldots,E_n)\geq0\,.
}
Thus open-universe positivity directly constrains the inverse Laplace transform of every multi-boundary amplitude.

Second, we can glue slices that connect their codimension-two boundaries in different ways \cite{Colafranceschi:2023txs, Colafranceschi_2024}. Consider four boundary components and two open intervals. Let $\cJ_A$ connect components $(1,2)$ and $(3,4)$, while $\cJ_B$ connects $(1,4)$ and $(2,3)$. For simplicity, take every interval to have length $\beta/2$. Gluing a slice to its own conjugate produces two circles of length $\beta$, whereas gluing the two inequivalent pairings produces one circle of length $2\beta$. The Gram matrix of $\ket{\cJ_A}$ and $\ket{\cJ_B}$ is therefore
\es{eq:pairing-gram-matrix}{
\begin{pmatrix}
G_2(\beta,\beta)&G_1(2\beta)\\
G_1(2\beta)&G_2(\beta,\beta)
\end{pmatrix}\succeq0\,.
}
In particular,
\es{eq:semi-quenched-positivity}{
G_2(\beta,\beta)\geq G_1(2\beta)\,.
}
Since $G_2(\beta,\beta)=\langle Z(\beta)^2\rangle$ and $G_1(2\beta)=\langle Z(2\beta)\rangle$, this is precisely the positivity of the second semi-quenched R\'enyi entropy~\cite{Antonini_2025, Antonini:2025rmr}. Under mild assumptions about the low-energy density of states~\cite{janssen2021lowtemperatureentropyjtgravity,okuyama2021quenchedfreeenergyspacetime,johnson2021quenchedfreeenergyjt}, imposing \eqref{eq:semi-quenched-positivity} at arbitrarily large $\beta$ requires every member of a factorizing ensemble to have an isolated ground state~\cite{Antonini_2025}.

\subsection{When the GPI computes a statistical average over boundary inner products}
\label{sec:open-universe-statistical-interpretation}

We now turn to the strongest interpretation considered in this paper. Suppose that the GPI computes a statistical average over boundary theories and that, in each realization, the amplitudes obtained by gluing open slices are themselves positive inner products. Schematically, this requires
\es{eq:open-statistical-interpretation}{
&G(J_1,\ldots, J_{ k})\\
&\quad=\int d\mu(\alpha)\,
\braket{\cJ_1}{\cJ'_1}_\alpha\cdots
\braket{\cJ_{ k}}{\cJ'_{ k}}_\alpha\\
&\quad=\int d\mu(\alpha)\,
Z_\alpha[ J_1]\cdots
Z_\alpha[ J_{ k}],
}
where $J_i=\cJ_i^*\cup\cJ'_i$. In every realization $\alpha$, one must have
\es{eq:realization-open-positivity}{
Z_\alpha(\cJ_a^*\cup\cJ_b)\succeq0\,.
}
Thus the random variables $Z_\alpha[J]$ cannot take arbitrary values.

For a finite collection of closed boundary conditions $J_1,\ldots,J_k$, define the set of random variables which satisfy the inner product positivity:
\es{eq:def-K}{
K_{\{J_i\}}
=\left\{(z_1,\ldots,z_k)\in\mathbb C^k:\;
z_i=Z_\alpha[J_i]
\text{ and \eqref{eq:realization-open-positivity} holds}
\right\}.
}
 For thermal boundaries, for example, $K$ is the space of tuples obtained from completely monotonic functions $Z_\alpha(\beta_i)$, together with the additional inequalities that follow from non-trivial open-universe gluings which impose that the energy spectrum that determines $Z_\alpha(\beta_i)$ is discrete. Together this means that $Z_\alpha(\beta_i)$ is the partition function of a system with discrete states with positive degeneracies but not necessarily integer~\cite{kaidi2020discretenessintegralityconformalfield,chiang2024geometrymodularbootstrap}. 

The statistical measure in \eqref{eq:open-statistical-interpretation} must be supported on $K$. The Riesz--Haviland criterion therefore becomes
\es{eq:stat-open}{
L_G[P]\geq0
\qquad\forall\,P\in\mC_{\rm open}^{\rm stat},
\qquad
\mC_{\rm open}^{\rm stat}
=\bigcup_{k\geq1}\operatorname{Pos}(K)\,.
}
Here $\operatorname{Pos}(K)$ denotes the polynomials that are non-negative when restricted to $K$. Such a polynomial can be negative at points in the ambient space $\mathbb C^k\setminus K$. Since $K\subsetneq\mathbb C^k$, the cone $\operatorname{Pos}(K)$ is larger than the cone of polynomials that are non-negative on all of $\mathbb C^k$. This means the open universe positivity constraints automatically imply the closed universe ones in the statistical interpretation. Geometrically this follows since an arbitrary closed manifold $J$ can always be decomposed into a (not necessarily symmetric) cut $J=\mathcal{J}_1^*\cup \mathcal{J}_2 $ with arbitrary choice of $\mathcal{J}_1^*,\mathcal{J}_2$. Then the GPI $G(J)$ can be written as an open inner product $G(J)=\braket{\mathcal{J}_1}{\mathcal{J}_2}$ and it follows that any $Z_{\alpha}(J)=\braket{\mathcal{J}_1}{\mathcal{J}_2}_{\alpha}$. Note that the cut is not on a fixed geometry, rather it is a cut on the boundary conditions which define the GPI and thus includes the sum over all geometries with boundary $J$. Since the choice of cut is arbitrary, the above is valid $\forall \mathcal{J}_1^*, \mathcal{J}_2 $ and different choices can produce independent constraints. 

In the open universe case, the distinction between SOS and non-SOS polynomials is not a useful criterion for distinguishing whether the  GPI computes inner products or statistical averages. We can indeed find positive polynomials $Q\in \mC_\text{open}^\text{stat}$ which are non-SOS and yet the positivity of the GPI $L_G(Q)\geq 0$ follows from positivity of the open universe GPI inner product, whose positivity cone we denote by $\mathcal C_\text{open}^\text{Hilbert}$. A simple example is: 
\es{eq:Q-open-polynomial}{
Q(z_1,z_2,z_3)=z_1+z_2-2z_3
}
which is indeed positive on the domain
\es{eq:K2-domain}{
K=\left\{(z_1,z_2,z_3)\in\mathbb R^3:\;
z_1,z_2\geq0,\quad z_3^2\leq z_1z_2\right\}.
}
To obtain the domain in \eqref{eq:K2-domain} from the GPI, take
\es{eq:z-beta-identification}{
z_1=Z_\alpha(\beta_1),\qquad
z_2=Z_\alpha(\beta_2),\qquad
z_3=Z_\alpha(\bar\beta),
}
where  $\bar\beta=\frac{\beta_1+\beta_2}{2}$.
Then \eqref{eq:K2-domain} is precisely the positivity of the Gram matrix of $\ket{\beta_1/2}_\alpha$ and $\ket{\beta_2/2}_\alpha$. Although $Q$ is not globally non-negative and cannot be an ordinary SOS polynomial, its averaged positivity already follows from the open-universe Hilbert-space condition:
\es{eq:Q-as-open-norm}{
L_G[Q]
=\left\lVert
\ket{\frac{\beta_1}{2}}
-\ket{\frac{\beta_2}{2}}
\right\rVert^2\geq0\,.
}
Thus a polynomial can be non-SOS in the ambient variables and still give no condition beyond open-universe norm positivity.

To isolate a genuinely statistical constraint that belongs to
Sec.~\ref{sec:open-universe-statistical-interpretation} but is not implied by
Sec.~\ref{sec:open-universe-conventional-interpretation}, we consider the
zeroth, first, and second moments of the density of states together.  Divide
the energy axis into four windows, labeled by $a=1,\ldots,4$, and let
\es{eq:rho-alpha-bins}{
\rho_{\alpha,a}\geq0
}
denote the density of states in an $\alpha$-eigenstate, smeared over the
$a$-th window.  Define
\begin{align}
\rho_{1,a}
&=\int \mathrm{d} \mu(\alpha)\,\rho_{\alpha,a},\nonumber\\
\rho_{2,ab}
&=\int \mathrm{d}\mu(\alpha)\,
\rho_{\alpha,a}\rho_{\alpha,b}.
\label{eq:rho2-bins}
\end{align}
We now introduce the Gram matrix obtained from inner products of the 5-component vector
${\boldsymbol v}(\alpha)=\left(1,\rho_{\alpha,1},\rho_{\alpha,2},\rho_{\alpha,3},\rho_{\alpha,4}\right)^T,$
\begin{align}
\mathcal{M}
&=\int \mathrm{d}\mu(\alpha)\,{\boldsymbol v}(\alpha)\, {\boldsymbol v}^T(\alpha)
 =
\begin{pmatrix}
1&\brho_1^T\\
\brho_1&{\mathbf{\rho}}_2
\end{pmatrix},
\label{eq:horn-augmented-moment-matrix}
\end{align}

Closed-universe Hilbert-space positivity applied to the identity and the four
one-universe states requires $\mathcal{M} \succeq0$.  Direct open-universe cuts require
all of its entries to be non-negative.  Thus conditions II.A and III.A imply
\begin{align}
\text{II.A:}\quad&\mathcal{M}\succeq0,\qquad
\text{III.A:}\quad \mathcal{M}_{IJ}\geq0,
\label{eq:horn-rho2-discrete-checks-main}
\end{align}
In other words, $\M$ belongs to the cone of doubly non-negative
$5\times5$ matrices.

A statistical interpretation is stronger.  Since every ${\boldsymbol v}$ is
entrywise non-negative, \eqref{eq:horn-augmented-moment-matrix} requires
$\M$ to live in a cone defined by\footnote{Without loss of generality, we may replace the integral by a sum with at most $d$ terms, where $d=\dim \bf x$.}:
\begin{align}
\M\in
\left\{
\sum_{r=1}^{d}\bx_r\bx_r^T:
(\bx_r)_I\geq0
\right\}.
\label{eq:CP-condition}
\end{align}
Matrices which satisfy this property are referred to as {\it completely positive} matrices~\cite{pfeffer2021cone5times5completely,Shaked_Monderer_2014}. They can be characterized by a dual constraint known as co-positivity: a matrix
$H$ is copositive if
$\bx^T H\bx\geq0$ for every entrywise non-negative vector $\bx$. For such matrices,  \eqref{eq:horn-augmented-moment-matrix}  immediately implies %
\begin{align}
\Tr(H \M)
=\int \mathrm d \mu(\alpha)\,{\boldsymbol v}(\alpha)^T H{\boldsymbol v}(\alpha)
\geq0.
\label{eq:horn-copositive-ensemble}
\end{align}
We can therefore find statistical constraints that are not generated by either type of Hilbert-space positivity by choosing a copositive matrix that is not a sum of a PSD matrix and a point-wise positive matrix. This is possible for sufficiently large matrices, $n \geq 5$. The simplest example of such a matrix is the Horn matrix~\cite{laurent2022exactnesssumofsquaresapproximationscone,nishijima2026copositivecompletelypositivecones},
\begin{equation}
H=\begin{pmatrix}
 1&-1& 1& 1&-1\\
-1& 1&-1& 1& 1\\
 1&-1& 1&-1& 1\\
 1& 1&-1& 1&-1\\
-1& 1& 1&-1& 1
\end{pmatrix}.
\label{eq:horn-matrix}
\end{equation}

One can check that for positive vectors $x_I \ge 0$, with $x_6=x_1$,
\begin{align}
\bx^T H \bx &=\left(\sum_{I=1}^5x_I\right)^2
 -4\sum_{I=1}^5x_I x_{I+1} \geq 0 %
 \label{eq:horn-copositive}
\end{align}
Thus \eqref{eq:horn-copositive-ensemble} is a necessary statistical
constraint involving amplitudes with at most two boundaries.

To emphasize that this condition is not implied by closed- and open-universe Hilbert-space positivity, we present an example in Appendix~\ref{sec:open-universe-Hilbert-space-positivity--without-statistical-positivity} for $\M$ that is PSD and point-wise positive, but violates \eqref{eq:horn-copositive-ensemble}. Just as in the closed universe case in sec. \ref{sec:toymodel}, the example of   $\M$  can be used to construct a topological toy model of gravity that satisfies the open-universe Hilbert space interpretation from section \ref{sec:open-universe-conventional-interpretation} but violates statistical positivity.

\section{Relation between the positivity cones}
\label{sec:positivity-cones}

Having defined the four sets of positivity constraints, it is instructive to summarize the relation between them through the following diagram:
\begin{equation}
\begin{tikzpicture}[
  cone/.style={font=\small},
  relation/.style={font=\small,inner sep=0pt}
]
  \def\sp{0.75}
  \node[cone] (open-stat) at (0,{\sp*2.6})
    {$\mC_{\rm open}^{\rm stat}$};
  \node[cone] (closed-stat) at ({-\sp*2.2},{\sp*1.3})
    {$\mC_{\rm closed}^{\rm stat}$};
  \node[cone] (open-Hilbert) at ({\sp*2.2},{\sp*1.3})
    {$\mC_{\rm open}^{\rm Hilbert}$};
  \node[cone] (closed-Hilbert) at (0,0)
    {$\mC_{\rm closed}^{\rm Hilbert}$};

  \node[relation,rotate=31]
    at ({-\sp*1.1},{\sp*1.95}) {$\subset$};
  \node[relation,rotate=149]
    at ({\sp*1.1},{\sp*1.95}) {$\subset$};
  \node[relation,rotate=149]
    at ({-\sp*1.1},{\sp*0.65}) {$\subset$};

  \begin{scope}[shift={({\sp*1.1},{\sp*0.65})},rotate=31]
    \draw[
      line width=0.65pt,
      dash pattern=on 0pt off 1.25pt,
      line cap=round
    ]
      (0.13,0.12) .. controls (-0.17,0.12) and (-0.17,-0.12) ..
      (0.13,-0.12);
  \end{scope}

  \node[font=\scriptsize]
    at ({\sp*1.5},{\sp*0.28}) {$(d>2)$};
\end{tikzpicture}
\label{eq:constraint-diagram}
\end{equation}
Here, $\mC_A \subset \mC_B$ implies that the set of constraints $\mC_B$ is {\it stronger} than $\mC_A$. If classifying gravitational EFTs into sets according to which of the constraints of \eqref{eq:constraint-diagram} they satisfy, the inclusion relations among those sets would be opposite to those in \eqref{eq:constraint-diagram}.

In sections \ref{sec:closed-universe-statistical-interpretation} and \ref{sec:open-universe-statistical-interpretation}, we have found that the statistical positivity constraints are stronger than the Hilbert space positivity constraints, due to the existence of polynomials whose positive expectation values are not implied by the positivity of any norm; this guarantees the inclusion relations going from right to left in the diagram above. 

The inclusion of $\mC_{\rm closed}^{\rm stat}$ in $\mC_{\rm open}^{\rm stat}$ is guaranteed by the fact that all $Z_\alpha[J]$ can be expressed as an inner product, $Z_\alpha[J] = \braket{\cJ}{\cJ'}_\alpha$ with $J = \cJ^* \cup \cJ'$. For $n$ possible choices of boundary conditions $J_i$, this constrains $Z_\alpha[J_i] $ to be part of $K \subset \mathbb C^n$; this thus makes the possible positivity constraints that we can impose on the moments of $Z_\alpha[J]$ in the open universe case much more restrictive. Note that, in the closed universe case, we did not restrict $Z_\alpha[J_i]$ in any way, and neither statistical nor Hilbert space positivity can constrain where $Z_\alpha[J_i]$ is sampled from. Thus, if we are solely interested in having a well-defined Hilbert space of closed universes but are not concerned about open universe positivity, we can, for instance, average over the partition function with sources,  $Z_\alpha[J]$,  of a non-unitary theory without a discrete spectrum. This is particularly relevant in cosmological settings, for example in de Sitter, where open universe positivity is not required in order to have well-defined transition probabilities between different states of the universe.

Finally the inclusion of  $\mC_{\rm closed}^{\rm Hilbert}$ in $\mC_{\rm open}^{\rm Hilbert}$ is more subtle: the condition $\braket{J_a}{J_b} \succeq 0 $ cannot directly be derived from $\braket{\cJ_a}{\cJ_b} \succeq 0$ by gluing open-universes with boundary conditions $\cJ_a^*$ and $\cJ_b$ to form closed universes with boundary conditions $J_a^*$ or $J_b$. This is, for instance, apparent when discussing boundaries with fixed ADM energies for which the two positivity conditions listed in \eqref{eq:horn-rho2-discrete-checks-main}  can clearly be satisfied independently. However, for $d>2$ spacetime dimensions one can instead find a choice of open universe boundary conditions $\cJ_a^*$ and $\cJ_b$ such that $\braket{\cJ_a}{\cJ_b} \succeq 0 $ implies $\braket{J_a}{J_b} \succeq 0$.\footnote{This is most clear in the AdS setting when $\cJ_a^*$, $\cJ_b$, $J_a^*$, and $J_b$ live on asymptotic AdS boundaries.} Specifically, starting from two closed universes with boundary conditions $J_a^*$ and $J_b$, one can eliminate an arbitrarily small $S^{d-2}$ portion of each of the $J_a^*$ and $J_b$ boundaries. Then, we can attach to each resulting open boundary an additional boundary with topology $I \times S^{d-2}$ where $I$ is an interval with flat metric and the size of $S^{d-2}$ is constant along the interval. Doing this leads to two open universes with boundary conditions $\cJ_a^*$ and $\cJ_b$ for which $\partial \cJ_a^*=\partial \cJ_b$ has topology $S^{d-2}$. In the limit when the proper length of each interval $I$ is very long, the geometry that dominates the inner-product $\braket{\cJ_a}{\cJ_b}$ connects the portion of the boundaries with boundary conditions $J_a^*$ and  $J_b$ by a closed universe, while the inside of $I \times S^{d-2}$  in $\cJ_a^*\cup\cJ_b$ is filled by the vacuum solution.\footnote{In 3d gravity, certain such manifolds and their pinching/degeneration limits  were considered in~\cite{Yin:2007at}. Pinching limits of 2D CFTs are well understood e.g.~\cite{Witten:2012bh,Gaberdiel:2010jf,Mason:2006dk}. } In this limit, we thus find that up to a multiplicative proportionality constant that is independent of the boundary conditions,
\begin{align}
\vcenter{\hbox{%
\begin{tikzpicture}[line cap=round]
  \definecolor{tubecutblue}{RGB}{31,119,180}
  \node[anchor=south west,inner sep=0] (tube) at (0,0)
    {\includegraphics[width=0.27\columnwidth]{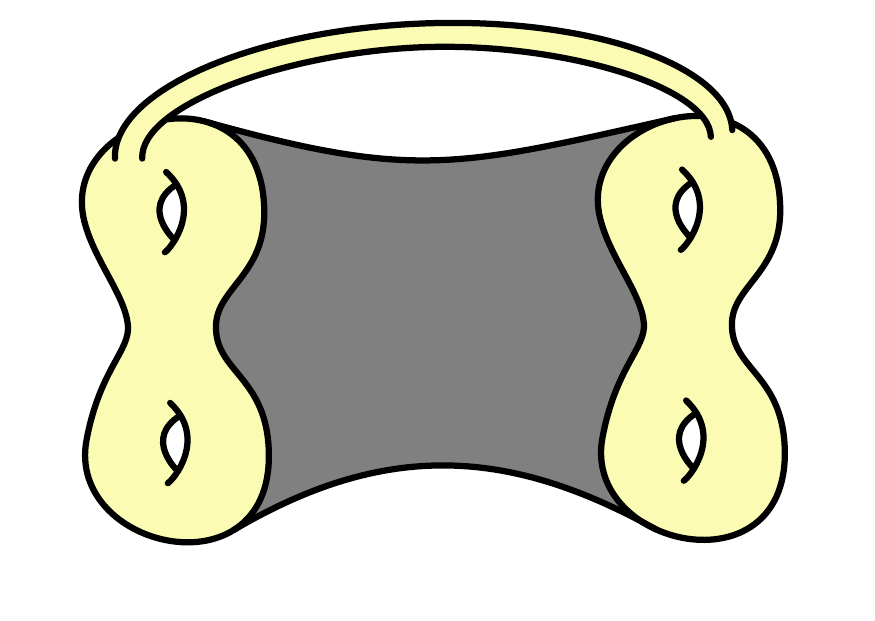}};
  \begin{scope}[x={(tube.south east)},y={(tube.north west)}]
    \path[draw=tubecutblue,line width=0.25pt,
          fill=tubecutblue,fill opacity=0.15]
      (0.505,1.044)
      .. controls (0.4795,1.044) and (0.4795,0.844) .. (0.505,0.844)
      .. controls (0.5305,0.844) and (0.5305,1.044) .. cycle;
    \draw[tubecutblue,line width=0.4pt,dash pattern=on 0.45pt off 0.35pt]
      (0.505,0.968) .. controls (0.499,0.968) and (0.499,0.920) .. (0.505,0.920);
    \draw[tubecutblue,line width=0.5pt]
      (0.505,0.920) .. controls (0.511,0.920) and (0.511,0.968) .. (0.505,0.968);
  \end{scope}
\end{tikzpicture}%
}}%
&\quad\,\,\sim\,\,\quad
\vcenter{\hbox{\includegraphics[width=0.23\columnwidth]{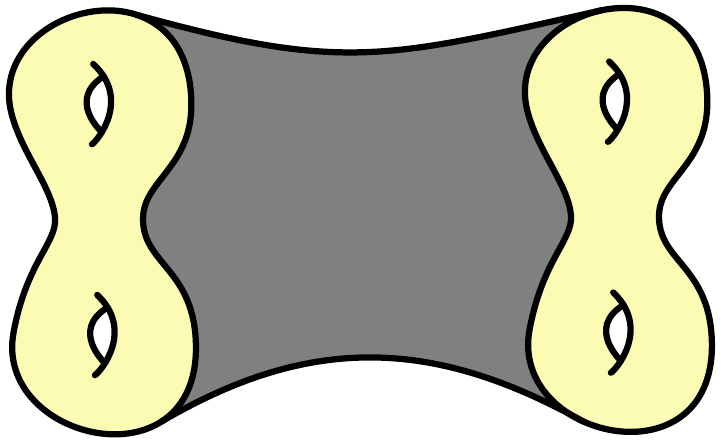}}}%
\label{eq:open-implies-closed-diagram}\\
\braket{\cJ_a}{\cJ_b}\succeq0
&\quad\Longrightarrow\quad \braket{J_a}{J_b}\succeq0
\end{align}

The figure on the right illustrates the case where $J_a^*$ and $J_b$ specify the boundary conditions on two genus two surfaces, whereas the figure on the left illustrates the example where $\cJ_a^*$ and $\cJ_b$ specify the boundary conditions on two genus two surfaces each with an $S^1$ boundary specified by the blue cut.\footnote{This is similar to the construction described in~\cite{Antonini:2023hdh,Antonini:2024mci,Antonini:2025ioh,Sasieta:2025vck}.}
Consequently, this construction implies that positive definiteness of the closed-universe inner product follows from positive definiteness of the open-universe inner product, explaining the bottom-right inclusion relation in \eqref{eq:constraint-diagram}.

\section{Discussion}
\label{sec:discussion}

In this paper, we distinguished two statements that are often treated as interchangeable. The first is that the GPI defines positive inner products between states of open and closed universes. The second is that its multi-boundary amplitudes are moments of a positive statistical ensemble.\footnote{A subset of the conditions necessary for interpretation (1) were also discussed in the context of large-$N$ filtering in~\cite{liu2025filteringcftslargen}. If the filter applied to products of observables that are positive for any $N$ preserves positivity, then interpretation (2) also holds. Consequently, one can always interpret the results of the filter as capturing moments in a statistical ensemble even if no explicit construction of the ensemble is available. We thank H.~Liu for discussions on this point.  } The first statement requires positivity on a cone generated by Hilbert-space norms. The second requires positivity on every polynomial that is non-negative on the support of the putative ensemble measure. These conditions agree in simple Gaussian approximations, where only single boundary geometries and two-boundary wormholes contribute to the GPI, but they are inequivalent in general.

For closed universes, the distinction is made through the known gap between non-negative and SOS polynomials. This leads to explicit constraints, such as \eqref{eq:motzkin-GPI-inequality} and \eqref{eq:choi-lam-GPI-inequality}, that involve higher-boundary wormhole amplitudes and are invisible to ordinary Gram-matrix positivity. For open universes, each realization in the ensemble must itself define a boundary Hilbert space. This restricts the support of the random variables to the domain $K$ and enlarges the set of positive test polynomials. The completely positive versus doubly non-negative matrix cones give the simplest manifestation of this distinction. The Horn inequality \eqref{eq:horn-copositive} is obeyed by every ensemble of positive densities of states but is not implied even after imposing both closed- and open-universe Hilbert-space positivity.

We presented a gravitational toy model makes the distinction between these positivity constraints concrete: violations of statistical positivity in the matter sector survive the sum over topologies leading to a violation of statistical positivity, while at the same time satisfying all Hilbert-space positivity conditions. Beyond this toy model, these constraints can be checked in theories for which sufficiently general wormhole amplitudes are known~\cite{Belin_2021,Collier_2022,de_Boer_2024,deboer2026surgerystatistics3dgravity,post2025nonrationalverlindeformulavirasoro,chandra2025statistics3dgravityknots}. They provide a way to classify low-energy effective theories according to which interpretation of the GPI they admit~\cite{mcnamara2026wormholesredherringsreflection,liu2025filteringcftslargen,Colafranceschi:2023txs}. The axion-wormhole analysis of our previous work was a first application of closed- and open-universe Hilbert-space positivity~\cite{DiUbaldo:2026rly}. Determining whether the same wormholes admit the stronger statistical interpretations requires control over amplitudes with more boundaries and with sufficiently varied boundary conditions. The inequalities developed here identify precisely which combinations of those amplitudes should be computed.

There is also an algebraic interpretation of the distinction between the different positivity conditions. This can be understood in the Marolf-Maxfield construction of the closed-universe Hilbert space, obtained through a GNS construction by acting on the no-boundary state $\ket{\emptyset}$ with a universe-creation operator $\widehat Z[J]$~\cite{Marolf:2020xie,gesteau2020holographicbabyuniversesobservable,Colafranceschi_2024,fewster2019algebraicquantumfieldtheory,Glimm:1987ylb}. Since GPI results are insensitive to the ordering of closed-universe boundary conditions (e.g., $G(J_1, J_2) = G(J_2, J_1)$), it then follows that $[\widehat Z[J_1],\widehat Z[J_2]]=0$. If the operators $\widehat Z(J)$ were bounded commuting normal operators, or unbounded self-adjoint operators whose spectral projections strongly commute, they would admit a simultaneous spectral decomposition~\cite{Reed:1975uy,putinar2008multivariatedeterminateness,schmüdgen2026domaincharacterizationsstrongcommutativitybreak}. Evaluating them in the Hartle--Hawking state would then automatically produce a positive joint measure, and all statistical positivity constraints would follow. Therefore, if the Hilbert-space conditions hold while a non-SOS statistical constraint fails, the commutation relation $[\widehat Z[J_1],\widehat Z[J_2]]=0$
cannot be promoted to such a joint spectral representation on the domain relevant to the GPI. This is possible for unbounded operators: commutativity on a common dense domain does not imply essential self-adjointness or strong commutativity of their spectral measures. Statistical positivity is precisely the additional condition needed for the moments of these operators to arise from a common positive spectral measure. %
To this end, going back to the gravitational example discussed in Section \ref{sec:toymodel} and Appendix \ref{sec:a-grav-example} is useful. Instead of writing the GPI over $ n$ boundaries as \eqref{eq:app-partitions}, we show that the GPI admits an equivalent decomposition into \textit{gravitational $\alpha$ sectors}: 
\es{}{
G_n(P_1,\ldots,P_n)
 =\sum_{\alpha=0}^\infty p_\alpha\,
 \langle\alpha,\Omega|\widehat Z(P_1)\cdots\widehat Z(P_n)|\alpha,\Omega\rangle,
}
where $\widehat Z(P)$ is a boundary creation operator with source $P$ and the gravitational $\alpha$-sectors are obtained by diagonalizing $\widehat Z(1)$, i.e., $\widehat Z(1) |\alpha,\Omega\rangle = \alpha |\alpha,\Omega\rangle$. This does not necessarily imply that they are also eigenstates of all $\widehat Z(P)$; in fact, the failure of non-SOS positivity for certain matter interactions proves that the operators $\widehat Z(P_i)$ do not strongly commute and therefore do not admit a joint spectral decomposition. Nevertheless, even if there does not exist a notion of $\alpha$-states for the GPI, the gravitational $\alpha$ sectors defined above provide a useful basis to construct the baby universe Hilbert space, which is no longer one-dimensional in a fixed gravitational $\alpha$ sector. To gain further intuition about operators that don't strongly commute, we provide a quantum mechanical example with commuting operators that are not strongly commuting in Appendix \ref{sec:what-fails-in-the-MM-construction}; the moments of such operators consequently violate the non-SOS positivity conditions presented above.

How powerful are the new constraints? One can get a sense of how powerful the statistical positivity constraints can be in comparison to the Hilbert space positivity constraints by bootstrapping a finite set of moments in a multivariate probability distribution or in a multi-matrix integral~\cite{lasserre2018momentsoshierarchy,keren2026stationarityenoughtightnessquantum,Han_2020,cho2025thermalbootstrapmatrixquantum,Pironio_2010,khalkhali2025bootstrappingcriticalbehaviormultimatrix}.\footnote{In a single-variable probability distribution or in a single-matrix integral, the statistical positivity constraints are implied by the Hilbert space positivity constraints.} We perform such an analysis in both settings in Appendix \ref{sec:non-SOS-bootstrap}. At fixed truncation, SOS positivity can be imposed by semidefinite programming, while non-SOS positivity can be approximated using hierarchies of copositive or moment-cone constraints~\cite{lasserre2018momentsoshierarchy,belin2024approximatecftsrandomtensor,klep2024sumssquarescertificatespolynomial}. Comparing the resulting bounds quantifies the power of the additional statistical positivity constraints: both for multivariate probability distributions and for multi-matrix integrals, we find that the constraints on the lowest moments improve significantly when imposing statistical positivity.

More broadly, the same logic applies to any statistical theory whose observables are computed as moments in a distribution with positive measure. This includes ordinary matrix integrals and statistical systems whose critical points might be described by either unitary (e.g., the Ising critical point) or non-unitary CFTs (e.g., the critical point of the Potts model), where positivity of the microscopic probability measure can survive even when reflection positivity of the continuum theory does not. We hope that the constraints described here initiate a broader program of imposing statistical positivity, beyond sums of squares, in gravitational path integrals, matrix models, and conformal field theories.

\section*{Acknowledgments}
We thank Daniel Harlow, Hong Liu, Douglas Stanford, and Zhenbin Yang for valuable discussions. We are especially thankful to Don Marolf for detailed comments. LVI and GDU were supported in part by the Leinweber Institute for Theoretical Physics
at UC Berkeley, by the Department of Energy, Office of Science, Office of High Energy
Physics through the award DE-SC0025522, and by the Department of
Energy through QuantISED award DE-SC0019380. LVI was supported by the DOE Early Career Award DE-SC0025522. GDU is supported by the Japan Science and Technology Agency (JST) as part of Adopting Sustainable Partnerships for Innovative Research Ecosystem (ASPIRE), Grant No.\ JPMJAP2318. CY was supported by the Simons Investigator in Physics Award MP-SIP-0001737 and U.S. Department of Energy grant DE-SC0007870.  This work was performed in part at the Aspen Center for Physics, which is supported by National Science Foundation grant PHY-2210452. We would also like to thank Google DeepMind for offering access to the DeepThink function, and OpenAI for offering the Pro academic subscription.

\appendix

\onecolumngrid
\section{How to generate an example of a pseudo-average }
\label{sec:pseudo-average-examples}

\subsection{Closed universe Hilbert space positivity without statistical positivity}
\label{sec:closed-universe-Hilbert-space-positivity--without-statistical-positivity}

The goal of this appendix is to provide a set of potential values for the GPI amplitudes $G_n(J_1, \dots, J_n)$ that satisfy Hilbert space positivity but violate statistical positivity, therefore defining a pseudoaverage. Towards that end, we define the Gram matrix of inner products between states with up to $n$-closed universes as $G^{(n)}$. $G^{(n)}$ takes the form of a block Hankel matrix, 
\es{}{
G^{(n)} = \left.G_{r+s}\right|_{r,s=0}^{n}=\begin{pmatrix}
G_0&G_1&\cdots&G_n\\
G_1&G_2&\cdots&G_{n+1}\\
\vdots&\vdots&\ddots&\vdots\\
G_n&G_{n+1}&\cdots&G_{2n}.
\end{pmatrix}
}
where $G_n$ is the block defined by all inner products involving $n$ closed universes.
Condition I.A
requires $G^{(n)}\succeq0$ for every $n$.
To construct an example of a pseudo-average we start at the two-universe level with 
\es{eq:G2}{
G^{(2)} = G^{(2)}_\text{(Gaussian)} + \Lambda \begin{pmatrix}
1&0&0\\
0&0&0\\
0&0&T
\end{pmatrix}\,,
}
where $G_{\rm Gaussian}^{(2)}$ is the moment matrix for monomials of degrees $0,1,2$ in four independent standard real
Gaussian variables.\footnote{The superscript $2$ labels
basis monomials of degree at most two; their pairings involve moments
through degree four.} The second term in \eqref{eq:G2} is a deformation of this Gaussian, where $T$ is the fully symmetric tensor with nonzero components
\begin{align}
T_{4444}&=\frac34,\qquad T_{1234}=1,\qquad
T_{1111}=T_{2222}=T_{3333}=2,\nonumber\\
T_{1122}&=T_{1133}=T_{1144}=T_{2233}=T_{2244}=T_{3344}=1,
\label{eq:closed-T-components}
\end{align}
including all index permutations, with all other entries equal to zero.

This defines the truncated functional $L_T$ by
\begin{equation}
L_T(1)=1,\qquad
L_T(x_{i_1}\cdots x_{i_m})=0\quad(1\leq m\leq3),
\qquad
L_T(x_ix_jx_kx_l)=T_{ijkl}.
\label{eq:LT-definition}
\end{equation}
The inner product between two closed universes is given by the $2|2$ flattening $T_{ij,kl}=\braket{i,j}{k,l}$. It is positive semidefinite $T_{ij,kl}\succeq0$, which we can see by  considering a general polynomial of degree at most two,
\begin{equation}
P(x)=c+\ell_i x_i+A_{ij}x_i x_j,
\qquad
A_{ij}=A_{ji},
\label{eq:closed-quadratic-test}
\end{equation}
and evaluating $L_{T}$ on its Hermitian square.  The definition
\eqref{eq:LT-definition} gives
\begin{align}
L_T(|P|^2)
={}&|c|^2+\overline A_{ij}T_{ij,kl}A_{kl}\nonumber\\
={}&|c|^2+\frac13\sum_{i<j\leq3}|A_{ii}-A_{jj}|^2
+\left|\tfrac{2}{\sqrt3}\sum_{i=1}^3A_{ii}+\tfrac{\sqrt3}{2}A_{44}\right|^2
+4\bigl(|A_{12}+A_{34}|^2+|A_{13}+A_{24}|^2+|A_{14}+A_{23}|^2\bigr)\geq0.
\label{eq:LT-square-positivity}
\end{align}
Thus \eqref{eq:LT-square-positivity} shows that this truncated functional is positive on every Hermitian square of
degree at most four (Hermitian-SOS positive).  Now consider the real non-SOS Choi--Lam quartic polynomial~\cite{ChoiLamReznick} evaluated on the Hermitian combinations $x_i=\operatorname{Re}z_i$:
\begin{equation}
p_{\rm CL}(x)
=
x_4^4+x_1^2x_2^2+x_1^2x_3^2+x_2^2x_3^2
-4x_1x_2x_3x_4.
\label{eq:appendix-choi-lam-polynomial}
\end{equation}
Equivalently,
$F_{\rm CL}(z,\bar z):=p_{\rm CL}(\operatorname{Re}z)\geq0$ is pointwise
nonnegative on $\mathbb C^4$.  Nevertheless, the Hermitian-SOS positive functional $L_T$ is negative on the Choi-Lam polynomial:
\begin{align}
L_T(p_{\rm CL})
&=T_{4444}+T_{1122}+T_{1133}+T_{2233}-4T_{1234}\\
&=\frac34+1+1+1-4=-\frac14.
\label{eq:LT-choi-lam-value}
\end{align}
We can then define a linear functional $L_{\Lambda}$ for the tensor $G^{(2)}$ in the same way as $L_T$ in \eqref{eq:LT-definition} but now including the Gaussian contribution from \eqref{eq:G2}.  The Gaussian
matrix is positive definite and the added block is positive
semidefinite, so $G_{\Lambda}^{(2)}\succeq0$ and satisfies II.A at the
two-universe level.  On the Choi--Lam polynomial, we find
\begin{equation}
L_{\Lambda}(p_{\rm CL})
=L_{\rm Gaussian}(p_{\rm CL})+\Lambda L_T(p_{\rm CL})
=6-\frac{\Lambda}{4}<0
\qquad(\Lambda>24).
\label{eq:closed-CL-violation}
\end{equation}
Thus $L_\Gamma$ violates the positivity condition II.B.  Its zeroth moment is $1+\Lambda$; if desired, one
can normalize the functional by dividing every moment by $1+\Lambda$,
which does not change the violation of positivity. 
Finally, we extend the example to arbitrary universe number. This can be done by induction. Suppose that
the Gram matrix with up to $2n-2$ universes is positive definite
$G^{(n-1)}\succeq0$. Consider the $2n$ universe Gram matrix and decompose it as: 
\begin{equation}
G^{(n)}=
\begin{pmatrix}
G^{(n-1)}&B_n\\
B^\dagger_n&G_{2n}
\end{pmatrix},
\qquad
B_n=\begin{pmatrix}G_n&\cdots&G_{2n-1}\end{pmatrix}^{T}.
\label{eq:closed-Gram-recursion}
\end{equation}
We will use that the Gram matrix is positive definite  $G^{(n)}\succ 0$ if and only if both $G^{(n-1)}\succ 0$ and its Schur complement $G^{(n)}/G^{(n-1)}\succ 0$ are positive definite.
Choose the new odd-order block to vanish $G_{2n-1}=0$.
Let $K_{2n}^{\rm Gaussian}$ be the degree-$2n$ Gaussian moment tensor and let 
$K_{n,n}^{\rm Gaussian}\succeq0$ be its $n|n$ flattening.  
Taking
$G_{2n}=c_nK_{2n}^{\rm Gaussian}$, the  Schur complement is positive definite provided that 
\begin{equation}
c_n > \lambda_{\max}\!\left[
(K_{n,n}^{\rm Gaussian})^{-1/2}
B_n^{\dagger}
(G^{(n-1)})^{-1}
B_n
(K_{n,n}^{\rm Gaussian})^{-1/2}
\right].
\label{eq:closed-Schur-bound}
\end{equation}
where $\lambda_{\max}$ denotes the largest eigenvalue. 
Iterating this step gives compatible positive-definite Gram matrices
$G^{(n)}$ for all $n$.  Since the recursion never changes the moments of
degree four or below, the negative value in
\eqref{eq:closed-CL-violation} remains.  The completed functional therefore
satisfies II.A at all orders while violating II.B and thus gives an example of a positive definite inner product for an arbitrary number of closed universes that does not admit a statistical interpretation.\\

We can give an example of explicit coefficients $c_n$ that are sufficient to ensure the completion to all orders. They can be chosen to be:
\begin{equation}
 R_n=B_n^\dagger\bigl(G^{(n-1)}\bigr)^{-1}B_n,
 \qquad
 c_n=1+\operatorname{Tr}\!\left[
 \bigl(K^{\mathrm{Gaussian}}_{n,n}\bigr)^{-1}R_n
 \right]
 \label{eq:app-cn-explicit}
\end{equation}
This is a sufficient, not necessarily minimal, choice of $c_n$. Indeed, the matrix
\begin{equation}
 Q_n=\bigl(K^{\mathrm{Gaussian}}_{n,n}\bigr)^{-1/2}
 R_n\bigl(K^{\mathrm{Gaussian}}_{n,n}\bigr)^{-1/2}
 \succeq0
 \label{eq:app-cn-positive-matrix}
\end{equation}
has maximal eigenvalue $\lambda_{\max}(Q_n)\leq\operatorname{Tr}Q_n=c_n-1$. Consequently, the condition $\lambda_{\max}(Q_n)\leq c_n-1<c_n$ is trivially satisfied and the completion 
\begin{equation}
 c_nK^{\mathrm{Gaussian}}_{n,n}-R_n
 =\bigl(K^{\mathrm{Gaussian}}_{n,n}\bigr)^{1/2}
 \bigl(c_n\mathbf 1-Q_n\bigr)
 \bigl(K^{\mathrm{Gaussian}}_{n,n}\bigr)^{1/2}
 \succeq K^{\mathrm{Gaussian}}_{n,n}\succ0.
 \label{eq:app-cn-schur}
\end{equation}
For vectors $u$ and $v$, respectively in the sectors of degree at most $n-1$ and
homogeneous degree $n$, completing the square gives
\begin{align}
 \begin{pmatrix}u\\v\end{pmatrix}^{\!\dagger}
 G^{(n)}
 \begin{pmatrix}u\\v\end{pmatrix}
 ={}&
 \left(u+\bigl(G^{(n-1)}\bigr)^{-1}B_nv\right)^\dagger
 G^{(n-1)}
 \left(u+\bigl(G^{(n-1)}\bigr)^{-1}B_nv\right)
 \nonumber\\
 &+v^\dagger
 \left(c_nK^{\mathrm{Gaussian}}_{n,n}-R_n\right)v>0
 \label{eq:app-cn-square}
\end{align}
for every $(u,v)\neq0$. Thus $G^{(n)}\succ0$, and iteration proves positivity for arbitrary finite universe number.

\subsection{Open universe Hilbert space positivity without statistical positivity}
\label{sec:open-universe-Hilbert-space-positivity--without-statistical-positivity}

This appendix supplies the explicit witness promised at the end of
Sec.~\ref{sec:open-universe-statistical-interpretation}.  At the two-universe level, the thermal
amplitude is the double Laplace transform of the energy-space density,
\begin{equation}
 G_2(\beta,\beta')
 =\int_0^\infty dE\,dE'\,
 \rho_2(E,E')e^{-\beta E-\beta'E'}.
 \label{eq:horn-rho2-transform}
\end{equation}
In the energy polarization, the relevant II.A and direct-open-cut III.A
tests are, respectively,
\begin{align}
 \text{II.A:}\quad &\rho_2\succeq0,\nonumber\\
 \text{direct-open-cut III.A:}\quad
 &\rho_2(E,E')\geq0.
 \label{eq:horn-rho2-two-checks}
\end{align}
For a finite-dimensional witness, choose five distinct energies
$E_a\geq0$.  The density becomes a $5\times5$ matrix and
\begin{equation}
 G_2(\beta,\beta')
 =\sum_{a,b=1}^5\rho_{2,ab}
 e^{-\beta E_a-\beta'E_b}.
 \label{eq:horn-G2-discrete}
\end{equation}
The two tests in \eqref{eq:horn-rho2-two-checks} reduce to
\begin{equation}
 \rho_2\succeq0,
 \qquad
 \rho_{2,ab}\geq0
 \quad (a,b=1,\ldots,5).
 \label{eq:horn-rho2-discrete-checks}
\end{equation}
These conditions would imply the Horn inequality for every $\rho_2$ only
if $H=P+N$, where $P\succeq0$ and $N$ is entrywise
nonnegative~\cite{laurent2022exactnesssumofsquaresapproximationscone}.
The Horn matrix \eqref{eq:horn-matrix} has no such decomposition.

Indeed, choose
\begin{equation}
 \rho_2=
 \begin{pmatrix}
  1&\frac35&0&0&\frac35\\
  \frac35&1&\frac35&0&0\\
  0&\frac35&1&\frac35&0\\
  0&0&\frac35&1&\frac35\\
  \frac35&0&0&\frac35&1
 \end{pmatrix}.
 \label{eq:horn-rho2-witness}
\end{equation}
It is entrywise nonnegative, and its eigenvalues are
$1+\frac65\cos(2\pi k/5)>0$ for $k=0,\ldots,4$, so it is positive
definite.  Nevertheless,
\begin{equation}
 \sum_{a,b=1}^5H_{ab}\rho_{2,ab}=-1.
 \label{eq:horn-separation}
\end{equation}
Thus \eqref{eq:horn-rho2-witness} passes the II.A and direct-open-cut III.A
tests but violates the necessary III.B constraint
\eqref{eq:horn-copositive-ensemble}.  This explicitly separates
inner-product positivity from the statistical interpretation already at the
two-universe level.\footnote{For a discrete microscopic spectrum, the full
two-point density also contains the equal-level contact term
$\overline\rho(E)\delta(E-E')$~\cite{Saad:2019lba}.  To treat it without
evaluating a distribution at coincident sharp energies, choose nonnegative,
disjoint, $L^2$-normalized windows $f_a(E)$ and define
$\rho_{2,ab}^{(f)}=\int dE\,dE'\,
f_a(E)\rho_2(E,E')f_b(E')$.  The contact term then becomes the finite
diagonal matrix
$C_{ab}=\delta_{ab}\int dE\,\overline\rho(E)f_a(E)^2$.
The displayed matrix may be understood as this full smeared density, with
the contact contribution already included on its diagonal.  Equivalently,
if it is displayed separately, then
$\langle H,\rho_2^{(f)}\rangle
=\langle H,\rho_2^{(f)}-C\rangle+\operatorname{Tr}C$; the violation remains
whenever the negative regular contribution has magnitude greater than
$\operatorname{Tr}C$.  Smearing makes the contact term finite and retains it.}

\subsection{A gravitational example}

\label{sec:a-grav-example}

In this section we construct an example of a gravitational theory for which the inner product between arbitrary closed universes is positive semidefinite but does not admit a statistical interpretation as it violates non-SOS positivity.
\\

\paragraph{The model. }As summarized in the main text, we start from the Marolf-Maxfield topological model and modify it by adding a model of interacting matter, whose properties are as follows. Let  $x_i$, $i=1,\ldots,r$ indicate $r$ real matter sources, which can be inserted on any boundary. In particular, we can think of the sources as operators $\widehat{O}_i$ acting on the boundary Hilbert space of the putative dual quantum system that mutually commute $[\widehat{O}_i,\widehat{O}_j] = 0$. Consider a surface with $n$ boundaries; a polynomial $P_k(\mathbf{x})$ in the sources specifies the insertions on boundary $k$.
We define a linear functional $\ell$ on polynomials which encodes the matter inner product. In boundary terms, 
\es{}{\ell(x_{i_1} \cdots x_{i_n}) = \ev{\widehat{O}_{i_1} \cdots  \widehat{O}_{i_n}}.}
The functional is chosen so that it is normalized and SOS-positive: 
\begin{equation}
 \ell(1)=1,\qquad \ell\left(P^*(\mathbf{x})P(\mathbf{x})\right)>0\quad(P(\mathbf{x})\ne0).
 \label{eq:app-state}
\end{equation}
SOS-positivity guarantees that the inner product for a reflection-symmetric matter configuration is positive. 

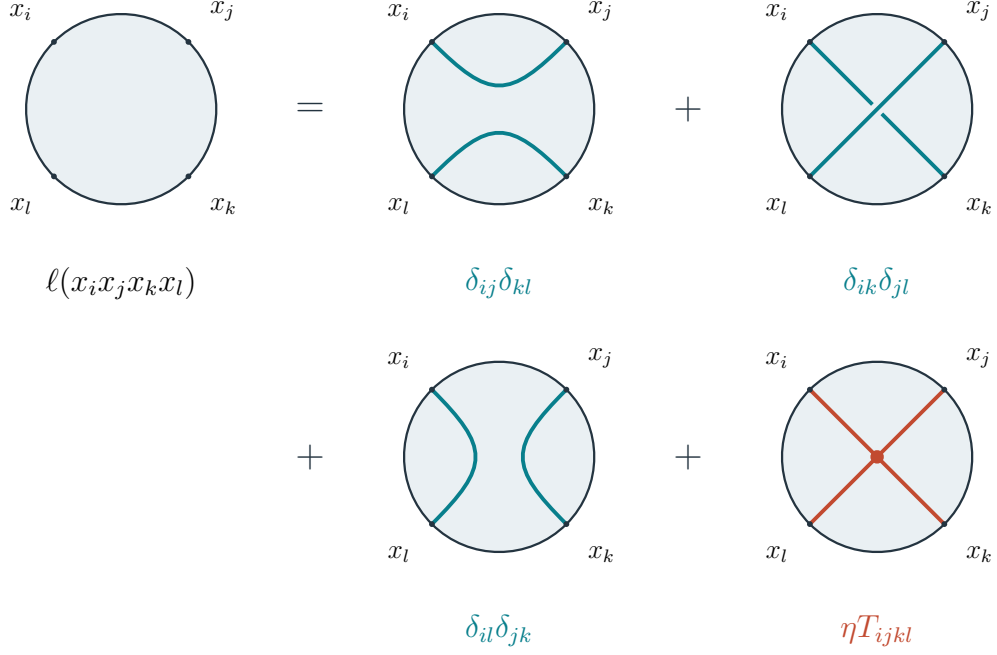
\begin{figure}[!t]
    \centering
\begingroup
\renewcommand{\sdMatterMarks}{%
 \foreach \x/\y in {0/1.48,1.48/0,0/-1.48,-1.48/0}
  \fill[sdink] (\x,\y) circle (.045);
 \node[above left,inner sep=2pt] at (0,1.85) {$x_i$};
 \node[above right,inner sep=2pt] at (1.85,0) {$x_j$};
 \node[below right,inner sep=2pt] at (0,-1.85) {$x_k$};
 \node[below left,inner sep=2pt] at (-1.85,0) {$x_l$};
}
\begin{tikzpicture}[x=1cm,y=1cm,font=\fontsize{12}{15}\selectfont]
 \ifsdCompact
  \path[use as bounding box] (0,4.2) rectangle (24,10.9);
 \foreach \x/\mode in {2.60/0,7.10/1,11.60/2,16.10/3,20.60/4}{
  \begin{scope}[shift={(\x,8.55)},rotate=45,transform shape,every node/.append style={rotate=-45}]\sdDisk{\mode}\end{scope}
 }
 \sdEquationSymbols\sdMomentTerms
 \else
 \path[use as bounding box] (0,0) rectangle (15,10.6);
 \foreach \x/\y/\mode in {2.3/8.4/0,7.3/8.4/1,12.3/8.4/2,7.3/3.8/3,12.3/3.8/4}{
  \begin{scope}[shift={(\x,\y)},scale=.85,rotate=45,transform shape,every node/.append style={rotate=-45}]
   \sdDisk{\mode}
  \end{scope}
 }
 \node[sd formula] at (4.8,8.4) {$=$};
 \foreach \x/\y in {9.8/8.4,4.8/3.8,9.8/3.8}
  \node[sd formula] at (\x,\y) {$+$};
 \node at (2.3,6.1) {$\ell(x_ix_jx_kx_l)$};
 \node[text=sdwick] at (7.3,6.1) {$\delta_{ij}\delta_{kl}$};
 \node[text=sdwick] at (12.3,6.1) {$\delta_{ik}\delta_{jl}$};
 \node[text=sdwick] at (7.3,1.5) {$\delta_{il}\delta_{jk}$};
 \node[text=sdvertex] at (12.3,1.5) {$\eta T_{ijkl}$};
 \fi
\end{tikzpicture}
\endgroup
\vspace{-1.2cm}
    \caption{The disk amplitude with four matter sources is obtained by summing over Wick contractions plus the quartic interaction $T$.}
    \label{fig:disk4point}
\end{figure}

To be concrete, let's discuss how $\ell$ fixes the lower point matter correlators.  The one-point and three-point functions are chosen to vanish: 
\es{}{
\ell(x_i)=\ell(x_ix_jx_k)=0,
}
while the  two-point and four-point functions are equal to:
\begin{align}
 \ell(x_ix_j)&=\delta_{ij},\nonumber\\
 \ell(x_ix_jx_kx_l)&=W_{ijkl}+\eta T_{ijkl},
 \qquad
 W_{ijkl}=\delta_{ij}\delta_{kl}+\delta_{ik}\delta_{jl}+\delta_{il}\delta_{jk}.
 \label{eq:app-seed}
\end{align}
 The interaction that fixes the four-point function should be thought of as an effective renormalized interaction, not a tree-level vertex.  The $W$ tensor corresponds to Wick contractions  and $T$ is a positive semidefinite tensor chosen such that the matter functional violates statistical positivity $\ell(p_{\rm non-SOS}(x))<0$. The disk four-point amplitude is shown in Figure \ref{fig:disk4point}.  Since the interaction $T$ is positive semidefinite, it can always be expanded as a sum over three-point OPE-like coefficients: 
\es{defTCC}{
\begin{gathered}
T_{ijkl}=\sum_{a=1}^K C_{ij}^aC_{kl}^a.\\[6pt]
\begin{tikzpicture}[baseline=-.5ex,x=1cm,y=1cm,font=\small]
 \foreach \c in {0,3.25}{
  \begin{scope}[shift={(\c,0)}]
   \draw[sd body,line width=.65pt] (0,0) circle (.85);
   \foreach \x/\y in {-.601/.601,.601/.601,.601/-.601,-.601/-.601}
    \fill[sdink] (\x,\y) circle (.025);
   \node[above left,inner sep=1.5pt] at (-.601,.601) {$x_i$};
   \node[above right,inner sep=1.5pt] at (.601,.601) {$x_j$};
   \node[below right,inner sep=1.5pt] at (.601,-.601) {$x_k$};
   \node[below left,inner sep=1.5pt] at (-.601,-.601) {$x_l$};
  \end{scope}
 }
 \draw[sd interaction,line width=1pt] (-.601,.601)--(.601,-.601);
 \draw[sd interaction,line width=1pt] (.601,.601)--(-.601,-.601);
 \fill[sdvertex] (0,0) circle (.055);
 \node at (1.35,0) {$=$};
 \node at (1.95,0) {$\displaystyle\sum_{a=1}^{K}$};
 \begin{scope}[shift={(3.25,0)}]
  \draw[sd interaction,line width=1pt] (-.601,.601)--(0,.30)--(.601,.601);
  \draw[sd interaction,line width=1pt] (-.601,-.601)--(0,-.30)--(.601,-.601);
  \draw[sd interaction,line width=1pt] (0,.30)--(0,-.30);
  \fill[sdvertex] (0,.30) circle (.055);
  \fill[sdvertex] (0,-.30) circle (.055);
  \node[right,inner sep=3pt] at (0,0) {$a$};
 \end{scope}
\end{tikzpicture}

\end{gathered}
}
Physically we have cut the four-point function and obtained an $s$-channel decomposition, that is manifestly positive definite. 
We can complete the linear functional $\ell$, defined on up to four boundary sources (degree four polynomials), into a positive definite functional on an arbitrary number of sources by the procedure explained in the previous section.

\begin{figure}[!t]
    \centering
\begingroup
\renewcommand{\sdMatterMarks}{%
 \foreach \x/\y in {0/1.48,1.48/0,0/-1.48,-1.48/0}
  \fill[sdink] (\x,\y) circle (.045);
 \node[above left,inner sep=2pt] at (0,1.85) {$x_i$};
 \node[above right,inner sep=2pt] at (1.85,0) {$x_j$};
 \node[below right,inner sep=2pt] at (0,-1.85) {$x_k$};
 \node[below left,inner sep=2pt] at (-1.85,0) {$x_l$};
}
\begin{tikzpicture}[x=1cm,y=1cm,font=\fontsize{12}{15}\selectfont]
 \ifsdCompact
  \path[use as bounding box] (0,4.2) rectangle (24,10.9);
 \foreach \x/\mode in {2.60/0,7.10/1,11.60/2,16.10/3,20.60/4}{
  \begin{scope}[shift={(\x,8.55)},rotate=45,transform shape,every node/.append style={rotate=-45}]\sdWormhole{\mode}\end{scope}
 }
 \sdEquationSymbols\sdMomentTerms
 \else
 \path[use as bounding box] (0,0) rectangle (15,10.6);
 \foreach \x/\y/\mode in {2.3/8.4/0,7.3/8.4/1,12.3/8.4/2,7.3/3.8/3,12.3/3.8/4}{
  \begin{scope}[shift={(\x,\y)},scale=.85,rotate=45,transform shape,every node/.append style={rotate=-45}]
   \sdWormhole{\mode}
  \end{scope}
 }
 \node[sd formula] at (4.8,8.4) {$=$};
 \foreach \x/\y in {9.8/8.4,4.8/3.8,9.8/3.8}
  \node[sd formula] at (\x,\y) {$+$};
 \node at (2.3,6.1) {$\ell(x_ix_jx_kx_l)$};
 \node[text=sdwick] at (7.3,6.1) {$\delta_{ij}\delta_{kl}$};
 \node[text=sdwick] at (12.3,6.1) {$\delta_{ik}\delta_{jl}$};
 \node[text=sdwick] at (7.3,1.5) {$\delta_{il}\delta_{jk}$};
 \node[text=sdvertex] at (12.3,1.5) {$\eta T_{ijkl}$};
 \fi
\end{tikzpicture}
\endgroup
\vspace{-1.2cm}
    \caption{The four-boundary wormhole amplitude with one matter source on each boundary is obtained by summing over Wick contractions plus the quartic interaction $T$.}
    \label{fig:wormhole4pt}
\end{figure}
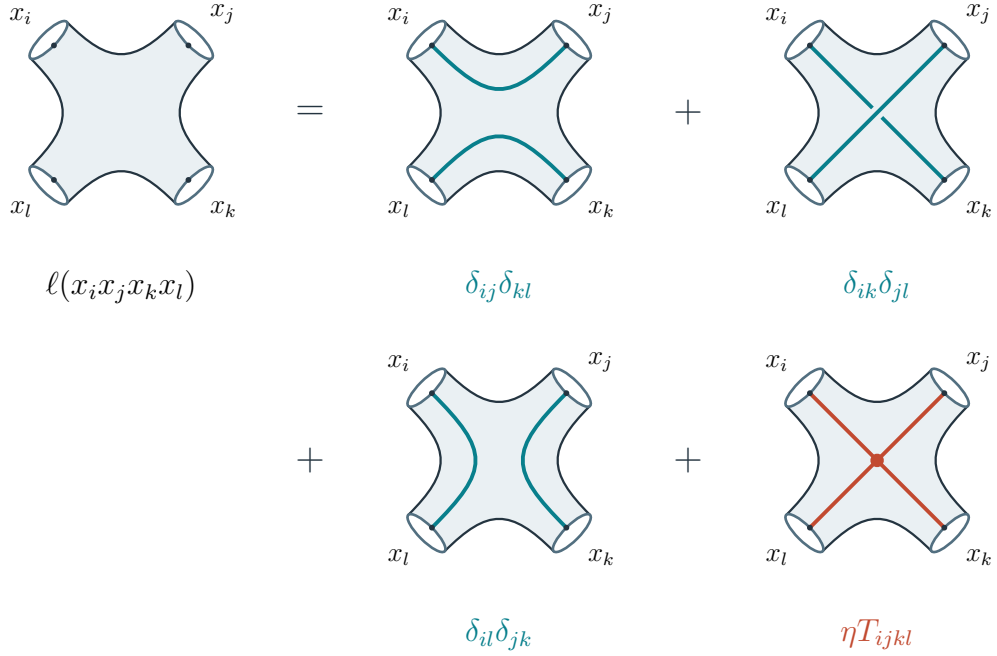

We now couple the topological matter to the Marolf-Maxfield gravitational model. The matter propagates only inside connected components of spacetime. A connected genus-$g$ surface with $n$  boundaries carrying sources $P_1,\ldots,P_n$ is assigned the weight\footnote{For $n=0$ the product is $1$.}
\begin{equation}
G_{g,n}^\text{conn}(P_1,\ldots,P_n)
 =e^{S_0(2-2g-n)}\,\ell(P_1\cdots P_n),\qquad S_0>0.
 \label{eq:appendix-connected}
\end{equation}

Note that the matter amplitude is the same independently of the number of boundaries, as the functional $\ell$ is evaluated on the product of the polynomials $P_1\dots P_n$. For example, a disk with four matter sources $P_{\rm disk}=x_1x_2x_3x_4$ (see Figure \ref{fig:disk4point}) and a four-boundary wormhole with a single source on each boundary $P_i=x_i$ (see Figure \ref{fig:wormhole4pt}) produce the same amplitude $\ell(P_{\rm disk})=\ell(P_1P_2P_3P_4)=\ell(x_1x_2x_3x_4)$. 
 To obtain the full GPI with specified boundaries and sources, we evaluate $G_{g,n}^\text{conn}$ for each connected component, summing over genus, then multiply the amplitudes of disconnected components, divide by factorials for identical compact components, and normalize by the no-boundary partition function. Summing over genus, a single connected-component gives a factor $\lambda$,
\begin{equation}
 \lambda=\sum_{g\ge0}e^{S_0(2-2g)}
 =\frac{e^{2S_0}}{1-e^{-2S_0}},\qquad Z_{\rm vac}=e^\lambda.
 \label{eq:app-lambda}
\end{equation}
The normalized $n$-boundary amplitude is therefore
\begin{equation}
 G_n(P_1,\ldots,P_n)
 =\sum_{\pi\in\Pi_n}\lambda^{|\pi|}
   \prod_{B\in\pi}\ell\!\left(\prod_{a\in B}P_a\right),
 \qquad G_0=1.
 \label{eq:appendix-partitions}
\end{equation}
The partition $\pi$ of the boundaries $(1,\dots,n)$ specifies which boundaries belong to the same connected spacetime, and $|\pi|$ is the number of connected components, weighted by $\lambda^{|\pi|}$. A block $B\in \pi$ is a connected spacetime for which the matter amplitude is given by multiplying all the sources in that block, irrespective of the boundary they belong to, hence $\ell\!\left(\prod_{a\in B}P_a\right)$. The outer product over $B$ is simply multiplying amplitudes for disconnected components. \\

\paragraph{A proof that inner products are positive definite.} 
Having computed the GPI for arbitrary boundaries and sources in this model, we proceed to prove that it realizes a positive definite inner product. To do so, we will construct the Hilbert space of baby universes with matter and rewrite $G_n$ as a manifestly positive definite inner product in this Hilbert space. The GPI without any matter sources $P_1=\dots=P_n=1$ reduces to the Marolf-Maxfield model, where the baby universe Hilbert space is given by a discrete set of $\alpha$ states denoted $\ket{\alpha}$ with $ \alpha=0,1,\dots \infty$ and the $n$ boundary amplitude is given by the $n$-th moment of a Poisson distribution in the $\alpha$ states: 
\es{eq:appendix-MM-GPI-ensemble}{
G_n(1,\dots,1) =\sum_{\pi\in\Pi_n}\lambda^{|\pi|}=\sum_{\alpha=0}^\infty p_\alpha \bra{\alpha}\widehat{Z}^n\ket{\alpha}=\sum_{\alpha=0}^\infty p_\alpha \alpha^n, \,\,\,\,\,\quad p_\alpha \equiv \frac{e^{-\lambda}\lambda^\alpha}{\alpha!}.
}
The value of $\alpha$ geometrically represents the number of (indistinguishable) connected components of spacetime weighted by  $\frac{\lambda^{\alpha}}{\alpha!}$ and the normalization factor $e^{-\lambda}$ is the norm of the no-boundary state. 

We proceed to construct the baby universe Hilbert space with matter by first working with a fixed number of connected components $\alpha$, constructing the Hilbert space for each connected component and successively summing over components $\alpha$ with measure $p_{\alpha}$. On an individual connected component, denote the vacuum state  as $\ket{\Omega}$ and a state with matter sources $P$ as $\ket{P}$, with the positive inner product   $\braket{P}{Q}_c=\ell(P^*Q)$. Define the multiplication operator $\mathsf M_P|Q\rangle=|PQ\rangle$. 
For $\alpha$ connected components the vacuum is simply the tensor product of the vacuum in each component which we denote as $\ket{\alpha,\Omega}=\ket{\alpha,\Omega_1,\dots,\Omega_\alpha}$ and a basis of product states is constructed by associating a polynomial $Q_k$ to each connected component $\ket{\alpha,Q_1,\dots,Q_\alpha}$. 
The Hilbert space is obtained by symmetrizing over the connected components $\mathcal H_\alpha=\operatorname{Sym}^\alpha\mathcal H_{\rm 1} $.
We construct the sourced boundary operators: 
\begin{equation}
\widehat{Z}(P)
=
\sum_{k=1}^{\alpha}
\mathbf 1^{\otimes(k-1)}
\otimes M_P
\otimes
\mathbf 1^{\otimes(\alpha-k)}=\sum_{k=1}^\alpha\mathsf M_P^{(k)} .
 \label{eq:app-dcopies}
\end{equation}

Its action on the Hilbert space of $\alpha$ connected components is to act by multiplication by a source $P$ on the $k$-th component and summing over $k$:
\begin{equation}
\widehat Z(P)
|\alpha,Q_1,\ldots,Q_\alpha\rangle
=
\sum_{k=1}^{\alpha}
|\alpha,Q_1,\ldots,PQ_k,\ldots,Q_\alpha\rangle .
\label{eq:Zd-polynomial-action}
\end{equation}
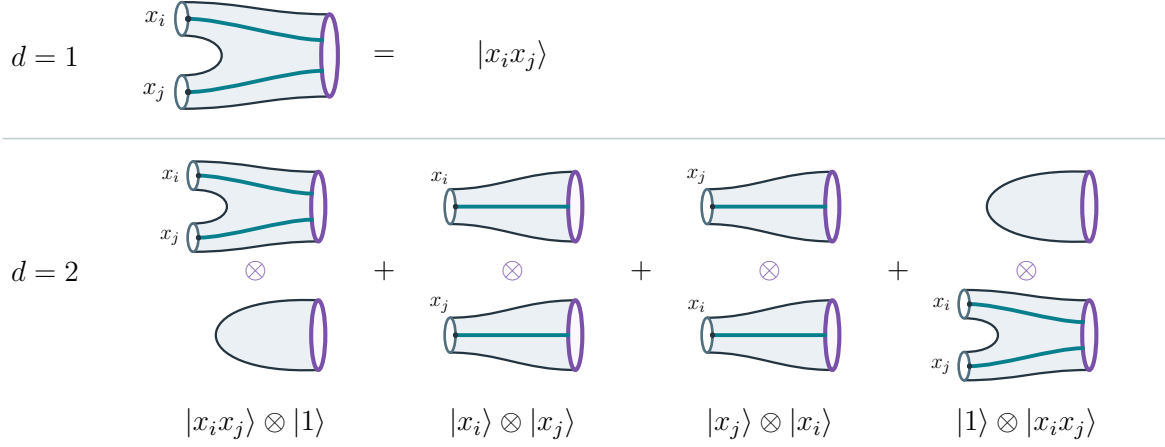
\begin{figure}[!t]
    \centering
    \begingroup
\let\ifsdProofReflected\iffalse
\let\ifsdProofWorldlines\iftrue
\newcommand{\sdSectorHalf}[3]{%
 \ifcase#1\relax
  \path[sd body] (1.15,.64)
   .. controls (-1.36,.75) and (-1.36,-.75) .. (1.15,-.64)--cycle;
 \or
  \path[sd body] (-1.15,.32)
   .. controls (-.25,.32) and (.20,.64) .. (1.15,.64)
   --(1.15,-.64)
   .. controls (.20,-.64) and (-.25,-.32) .. (-1.15,-.32)--cycle;
  \draw[sd rim] (-1.15,0) ellipse (.10 and .32);
  \fill[sdink] (-1.05,0) circle (.048);
  \node[above,inner sep=2pt] at (-1.34,.34) {$x_{#2}$};
 \or
  \path[sd body] (-1.15,.83)
   .. controls (-.15,.83) and (.16,.64) .. (1.15,.64)
   --(1.15,-.64)
   .. controls (.16,-.64) and (-.15,-.83) .. (-1.15,-.83)
   --(-1.15,-.31)
   .. controls (-.32,-.31) and (-.32,.31) .. (-1.15,.31)--cycle;
  \foreach \y in {.57,-.57}{
   \draw[sd rim] (-1.15,\y) ellipse (.10 and .26);
   \fill[sdink] (-1.05,\y) circle (.048);
  }
  \ifsdProofReflected
   \node[right,inner sep=3pt] at (-1.28,.57) {$x_{#2}$};
   \node[right,inner sep=3pt] at (-1.28,-.57) {$x_{#3}$};
  \else
   \node[left,inner sep=3pt] at (-1.28,.57) {$x_{#2}$};
   \node[left,inner sep=3pt] at (-1.28,-.57) {$x_{#3}$};
  \fi
 \fi
 \draw[sd cut,fill=sdcut!4] (1.15,0) ellipse (.13 and .64);
 \ifsdProofWorldlines
  \ifcase#1\relax
  \or
   \draw[sd wick] (-1.05,0)--(1.02,0);
   \fill[sdink] (-1.05,0) circle (.048);
   \fill[sdwick] (1.02,0) circle (.036);
  \or
   \foreach \sgn in {-1,1}{
    \draw[sd wick] (-1.05,{.57*\sgn})
     .. controls (-.50,{.57*\sgn}) and (.50,{.24*\sgn})
     .. (1.029,{.24*\sgn});
    \fill[sdink] (-1.05,{.57*\sgn}) circle (.048);
    \fill[sdwick] (1.029,{.24*\sgn}) circle (.036);
   }
  \fi
 \fi
}

\begin{tikzpicture}[x=1cm,y=1cm,font=\fontsize{11}{14}\selectfont]
 \path[use as bounding box] (0,0) rectangle (16,7.65);
 \node[anchor=west] at (.25,6.65) {$d=1$};
 \begin{scope}[shift={(3.6,6.65)},scale=.85,transform shape]
  \sdSectorHalf{2}{i}{j}
 \end{scope}
 \node at (5.3,6.65) {$=$};
 \node at (7,6.65) {$|x_ix_j\rangle$};
 \draw[draw=sdrim!35,line width=.6pt] (.25,5.55)--(15.65,5.55);

 \node[anchor=west] at (.25,3.8) {$d=2$};
 \foreach \x/\top/\bot/\u/\v in {3.6/2/0/i/j,7/1/1/i/j,10.4/1/1/j/i,13.8/0/2/i/j}{
  \begin{scope}[shift={(\x,4.65)},scale=.72,transform shape]
   \sdSectorHalf{\top}{\u}{\v}
  \end{scope}
  \begin{scope}[shift={(\x,2.95)},scale=.72,transform shape]
   \ifnum\bot=1
    \sdSectorHalf{1}{\v}{}
   \else
    \sdSectorHalf{\bot}{\u}{\v}
   \fi
  \end{scope}
  \node[text=sdcut] at (\x,3.8) {$\otimes$};
 }
 \foreach \x in {5.3,8.7,12.1}
  \node at (\x,3.8) {$+$};
 \node at (3.6,1.78) {$|x_ix_j\rangle\otimes|1\rangle$};
 \node at (7,1.78) {$|x_i\rangle\otimes|x_j\rangle$};
 \node at (10.4,1.78) {$|x_j\rangle\otimes|x_i\rangle$};
 \node at (13.8,1.78) {$|1\rangle\otimes|x_ix_j\rangle$};
\end{tikzpicture}
\endgroup
    \vspace{-1.2cm}
    \caption{Two matter insertions on two separate boundaries assigned to $d=1$ and $d=2$ connected components. The state $\ket{1} = \ket{\Omega}$, the Hartle-Hawking state.}
    \label{fig:matter-assignment-sectors}
\end{figure}
We represent the construction of such states in Figure \ref{fig:matter-assignment-sectors} with the calculation of overlaps between such states represented in Figure \ref{fig:six-boundary-overlap}.

In the sourceless $P=1$ case, the operator $\hat Z(P)$ reduces to the usual Marolf-Maxfield operator  $\widehat Z(1)=\widehat{Z}=\alpha\,\mathbf 1$ which counts the number of connected components of spacetime. Its eigenstates are states with fixed number of connected components $\alpha$. More specifically, the states $\ket{\alpha,Q_1,\dots,Q_\alpha}$ for arbitrary $Q_i$ are all eigenstates of $\widehat{Z}(1)$ since they have the same number of connected components. As a consequence, for fixed $\alpha$ the baby-universe Hilbert space is spanned by the polynomials $Q_k$. The choice to construct the Hilbert space starting with a fixed number $\alpha$ of connected components is equivalent to diagonalizing $\widehat Z(1)$ and working in its eigenbasis where the operators $\widehat{Z}(P)$ are generally not diagonal.

We now claim that the GPI for the Marolf-Maxfield model coupled to topological matter in (\ref{eq:appendix-partitions}) is equivalent to:
\begin{equation}
 G_n(P_1,\ldots,P_n)
 =\sum_{\alpha=0}^\infty p_\alpha\,
 \langle\alpha,\Omega|\widehat Z(P_1)\cdots\widehat Z(P_n)|\alpha,\Omega\rangle.
 \label{eq:app-hilbert}
\end{equation}
Using this representation we can prove the GPI is manifestly positive definite. 
Consider an arbitrary collection of sources $\mathbf P_A=(P_{A,1},\ldots,P_{A,n_A})$ on any number of boundaries $n_A$ and consider an arbitrary state given by a linear combination $\sum_{A}c_A \mathbf P_A $.  Its norm is then manifestly positive: 
\begin{equation}
\displaystyle
 \sum_{A,B}\overline c_Ac_B\,
 G(\mathbf P_A^*,\mathbf P_B)
 =\sum_{\alpha=0}^\infty p_\alpha
 \left\|\sum_Ac_A\prod_{j=1}^{n_A} \widehat Z(P_{A,j})
 |\alpha,\Omega\rangle\right\|^2\ge0.
 \label{eq:app-master}
\end{equation}
Every term is the norm of a state in the tensor product Hilbert space with positive inner-product. For any collection of sources $\mathbf P_A$ on an arbitrary number of boundaries, the state inside the norm contains at most polynomially many terms in $\alpha$, each of finite norm. The Poisson sum over $\alpha$ therefore converges.

The remaining step is to prove the Hilbert space representation (\ref{eq:app-hilbert}) is equal to the geometric expression (\ref{eq:appendix-partitions}). We start by expanding the operators $\widehat Z(P)=\sum_{r=1}^\alpha \mathsf M_P^{(r)}$ in (\ref{eq:app-hilbert}): 
\es{}{
\langle\alpha,\Omega|\prod_{a=1}^n\widehat Z(P_a)|\alpha,\Omega\rangle= \sum_{r_1,\dots,r_n}^{\alpha} \prod_{r=1}^\alpha\ell\left(\prod_{a: \,r_a=r}P_a\right)
}
This follows from the tensor product structure, insertions of sources to the same component are multiplied together and evaluated by $\ell$ while different copies factorize. The list $(r_1,\dots,r_n)$ determines a partition $\pi$ where boundaries are in the same block $B\in \pi$ when they belong to the same connected component, meaning to the same copy inside the tensor product.  The possible ways to assign $|\pi|$ blocks  to $\alpha$ distinct copies is the falling factorial $(\alpha)_{|\pi|}=\frac{\alpha!}{(\alpha-|\pi|)!}$. We then have: 
\es{}{
\langle\alpha,\Omega|\prod_{a=1}^n\widehat Z(P_a)|\alpha,\Omega\rangle=\sum_{\pi \in \Pi_n}(\alpha)_{|\pi|}\prod_{B \in \pi }\ell \left( \prod_{a\in B}P_a\right).
}
We can now sum over $\alpha$ with the Poisson distribution, as the only $\alpha$ dependence is in the first factor: 
\es{}{
\sum_{\alpha=0}^{\infty}p_{\alpha}(\alpha)_{|\pi|}=e^{-\lambda}\sum_{\alpha=|\pi|}^{\infty}\frac{\lambda^\alpha}{\alpha!}\frac{\alpha!}{(\alpha-|\pi|)!}=\lambda^{|\pi|}
}
where we used that $\frac{\alpha!}{(\alpha-|\pi|)!}=0$ for $\alpha<|\pi|$. Substituting this back, we obtain the equivalence with the GPI (\ref{eq:appendix-partitions}) which completes the proof of positive definiteness.\\

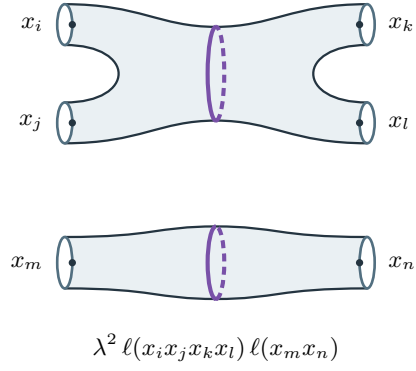
\begin{figure}[!t]
    \centering
    \begin{tikzpicture}[x=1cm,y=1cm,font=\small]
 \begin{scope}[yshift=2.5cm]
  \path[sd body]
   (-2,.92) .. controls (-1,.92) and (-.75,.62) .. (0,.62)
   .. controls (.75,.62) and (1,.92) .. (2,.92)
   --(2,.38) .. controls (1.05,.38) and (1.05,-.38) .. (2,-.38)
   --(2,-.92) .. controls (1,-.92) and (.75,-.62) .. (0,-.62)
   .. controls (-.75,-.62) and (-1,-.92) .. (-2,-.92)
   --(-2,-.38) .. controls (-1.05,-.38) and (-1.05,.38) .. (-2,.38)
   --cycle;
  \foreach \x in {-2,2}
   \foreach \y in {-.65,.65}{
    \draw[sd rim] (\x,\y) ellipse (.10 and .27);
   }
  \foreach \s in {-1,1}
   \foreach \t in {-1,1}{
    \fill[sdink] ({1.9*\s},{.65*\t}) circle (.048);
   }
  \node[left,inner sep=5pt] at (-2.1,.65) {$x_i$};
  \node[left,inner sep=5pt] at (-2.1,-.65) {$x_j$};
  \node[right,inner sep=5pt] at (2.1,.65) {$x_k$};
  \node[right,inner sep=5pt] at (2.1,-.65) {$x_l$};
  \draw[sd cut,densely dashed] (0,-.62)
   arc[start angle=-90,end angle=90,x radius=.11,y radius=.62];
  \draw[sd cut] (0,.62)
   arc[start angle=90,end angle=270,x radius=.11,y radius=.62];
 \end{scope}
 \path[sd body]
  (-2,.34) .. controls (-.8,.34) and (-.8,.48) .. (0,.48)
  .. controls (.8,.48) and (.8,.34) .. (2,.34)
  --(2,-.34) .. controls (.8,-.34) and (.8,-.48) .. (0,-.48)
  .. controls (-.8,-.48) and (-.8,-.34) .. (-2,-.34)--cycle;
 \foreach \x in {-2,2}{
  \draw[sd rim] (\x,0) ellipse (.10 and .34);
 }
 \foreach \s in {-1,1}{
  \fill[sdink] ({1.9*\s},0) circle (.048);
 }
 \node[left,inner sep=5pt] at (-2.1,0) {$x_m$};
 \node[right,inner sep=5pt] at (2.1,0) {$x_n$};
 \draw[sd cut,densely dashed] (0,-.48)
  arc[start angle=-90,end angle=90,x radius=.11,y radius=.48];
 \draw[sd cut] (0,.48)
  arc[start angle=90,end angle=270,x radius=.11,y radius=.48];
 \node at (0,-1.12)
  {$\displaystyle\lambda^2\,\ell(x_ix_jx_kx_l)\,\ell(x_mx_n)$};
\end{tikzpicture}
    \caption{A contribution to the overlap of two three-boundary states $\left(\bra{x_i x_j}\otimes \bra{x_m}\right) \left(\ket{x_k x_l} \otimes \ket{x_n} \right)$. Purple curves mark the cuts. Note that there is no $1/(2!)$ factor; if we label the baby universes as distinguishable, we would include a $1/(2!)$ but also we would sum over the $2!$ rearrangements. %
    }
    \label{fig:six-boundary-overlap}
\end{figure}

\paragraph{Failure of statistical positivity for an explicit choice of matter interactions. } 
We can now give an explicit example of the failure of statistical positivity by choosing the tensor $T$ to be the one defined in appendix \ref{sec:closed-universe-Hilbert-space-positivity--without-statistical-positivity}, which is negative on the Choi-Lam non-SOS polynomial. We evaluate the GPI on the Choi-Lam polynomial $p_{CL}$ in the operators $\widehat{Z}(x_i)$ that create a closed universe with a matter source $x_i$, for $i=1,2,3,4$. The Choi-Lam polynomial is a linear combination of gravitational amplitudes with four boundaries $G_4(x_i,x_j,x_k,x_l)$, each with a source $x_i$:
\begin{equation}
 G_4(x_i,x_j,x_k,x_l)
 =\lambda\eta T_{ijkl}+\lambda(1+\lambda)W_{ijkl}.
 \label{eq:app-four}
\end{equation}
The first term is the contribution of a single connected $\mathcal{O}(\lambda)$ four-boundary wormhole, while the second includes both the Wick contractions on the four-boundary wormhole $\mathcal{O}(\lambda)$ and the Wick contractions on two disconnected cylinders $\mathcal{O}(\lambda^2)$.  The resulting expectation value for the Choi-Lam polynomial is negative for sufficiently large coupling $\eta$:
\begin{equation}
\displaystyle
 L_G[p_{\rm CL}(Z_1,Z_2,Z_3,Z_4)]
 =\lambda\left(6-\frac\eta4\right)+6\lambda^2<0
 \quad \quad\text{when}\quad\quad  \eta>24(1+\lambda).
 \label{eq:app-negative}
\end{equation}
The violation of non-SOS positivity shows that the gravitational theory obtained by coupling the Marolf--Maxfield model to the topological matter sector does not admit a statistical interpretation. The sourceless gravitational sector, by itself, still admits the usual $\alpha$-state decomposition with positive weights $p_\alpha$. What happens when including  matter sources is that the  boundary operators $\widehat{Z}(P)$ can no longer be represented simultaneously as classical random variables.

Nevertheless, the full gravitational path integral with arbitrary matter sources retains a useful and suggestive decomposition over the \emph{gravitational $\alpha$-sectors}. In particular, as shown in Eq.~(\ref{eq:app-hilbert}),
\begin{equation}
G_n(P_1,\ldots,P_n)
=
\sum_\alpha p_\alpha\,
\langle\alpha,\Omega|
\widehat Z_\alpha(P_1)\cdots
\widehat Z_\alpha(P_n)
|\alpha,\Omega\rangle .
\label{eq:grav-alpha-decomposition}
\end{equation}
This representation should not be confused with a genuine $\alpha$-state decomposition of the full gravitational theory with sources. The operators $\widehat Z_\alpha(P)$ cannot, in general, be jointly diagonalized; equivalently, although they commute algebraically on the common domain relevant to the GPI, they do not strongly commute (see Appendix \ref{sec:what-fails-in-the-MM-construction}). The decomposition above is instead obtained by diagonalizing only the sourceless gravitational operator,
\begin{equation}
\widehat Z(1)\,|\alpha,\Omega\rangle
=
Z_\alpha(1)\,|\alpha,\Omega\rangle ,
\label{eq:sourceless-alpha}
\end{equation}
and then evaluating the sourced operators $\widehat Z_\alpha(P)$ within each such gravitational $\alpha$-sector. In this sense, the theory admits an average over gravitational $\alpha$-sectors even though it does not admit a statistical interpretation for the full GPI with sources.

This structure suggests a generalization to more general gravitational theories. Consider a gravitational theory $G$, which admits a statistical interpretation with $\alpha$-states $|\alpha,\Omega\rangle$ and positive measure $d\mu(\alpha)$, and couple it to a reflection-positive but non-statistical matter theory. It is within reason that the deformed GPI $\tilde{G}$ admits an analogous decomposition into gravitational $\alpha$-states:
\begin{equation}
\widetilde G_n\!\left((J_1,P_1),\ldots,(J_n,P_n)\right)
=
\int d\mu(\alpha)\,
\langle\alpha,\Omega|
\widehat Z_\alpha(J_1,P_1)\cdots
\widehat Z_\alpha(J_n,P_n)
|\alpha,\Omega\rangle ,
\label{eq:general-grav-alpha-deformation}
\end{equation}
where $J$ denotes the gravitational boundary data and $P$ the matter source. Setting all matter sources to the identity recovers the original gravitational theory $G$,
\begin{equation}
\widetilde G_n\!\left((J_1,1),\ldots,(J_n,1)\right)
=
G_n(J_1,\ldots,J_n).
\label{eq:recover-gravity}
\end{equation}
The theory $\tilde{G}$ will not admit a statistical interpretation, and the operators $\widehat Z_\alpha(J,P)$ within a fixed gravitational $\alpha$-sector are not simultaneously diagonalizable. Accordingly, the baby-universe Hilbert space is not one-dimensional, even in a fixed gravitational $\alpha$-sector. It would therefore be extremely interesting to determine whether such a decomposition arises in more general theories such as JT gravity coupled to matter. In particular, the gravitational $\alpha$ sectors do not always have a geometrical interpretation as the number of connected components of spacetime; in JT gravity they can be understood as individual realizations of the dual random matrix model.\\

\paragraph{A proof that the failure of statistical positivity persists. }
For the Marolf--Maxfield coupling defined in \eqref{eq:appendix-partitions},
the condition $\eta>24(1+\lambda)$ in \eqref{eq:app-negative} is needed
to make the four-boundary Choi--Lam combination negative. This is unsatisfactory since it forces us to make the interactions among the matter fields extremely large. We now show that a failure of statistical positivity exists even for couplings $\eta=\mathcal{O}(1)$, by examining amplitudes with a higher number of boundaries.
More generally, we show that for any  matter functional $\ell$ that is negative on a non-SOS polynomial, the violation of statistical positivity persists after coupling it to the gravitational model. The key is to use additional sourceless boundaries $Z(1)$ to isolate the gravitational
$\alpha=1$ sector, where the sourced correlators reduce to those of $\ell$.

Let $p_\text{non-SOS}(x_1,\dots,x_r)$ be a  non-SOS positive polynomial of degree $m$ in $r$ variables such that 
\begin{equation}
 \ell(p_\text{non-SOS})<0.
 \label{eq:app-persistence-matter}
\end{equation}
As in the main text, $p_\text{non-SOS}(Z(x_1),\dots,Z(x_r))\equiv p_\text{non-SOS}(\mathbf Z)$ denotes the polynomial
obtained by replacing each $x_i$ by $Z(x_i)$ and $\mathbf Z=(Z(x_1),\dots,Z(x_r))$. Its monomials specify
products of boundaries, rather than the single boundary
$Z[p_\text{non-SOS}(\mathbf x)]$. Given an integer $M\geq2$, construct the polynomial
\begin{equation}
 s_M(x)=\left[x\prod_{j=2}^{M}\frac{j-x}{j-1}\right]^2.
 \label{eq:app-persistence-filter}
\end{equation}
This polynomial is nonnegative for real $x$ and satisfies
\begin{equation}
 s_M(1)=1,
 \qquad
 s_M(\alpha)=0\quad(\alpha=0,2,\ldots,M).
 \label{eq:app-persistence-zeros}
\end{equation}
Consequently, the polynomial in $r+1$ variables $s_M(Z(1))p_\text{non-SOS}(\mathbf Z)$ is pointwise nonnegative.
It is also non-SOS: setting $Z(1)= 1$ in an SOS decomposition would give
an SOS decomposition of $p_\text{non-SOS}$, contradicting
\eqref{eq:app-state} and $\ell(p_\text{non-SOS})<0$.

Since $\widehat Z(1)=\alpha\,\mathbf1$, the representation
\eqref{eq:app-hilbert} gives
\begin{align}
 L_G\!\left[s_M(Z(1))p_\text{non-SOS}(\mathbf Z)\right]
 ={}&\lambda e^{-\lambda}\ell(p_\text{non-SOS}(\mathbf{x}))
 \nonumber\\
 &+\sum_{\alpha>M}p_\alpha s_M(\alpha)
 \langle\alpha,\Omega|
 p_\text{non-SOS}\bigl(\widehat Z(x_1),\ldots,
                       \widehat Z(x_r)\bigr)
 |\alpha,\Omega\rangle.
 \label{eq:app-persistence-tail}
\end{align}
The fixed-$\alpha$ expectation in the second line is a polynomial in
$\alpha$ of degree at most $m$. Indeed, the partition counting used
above expresses a correlator with $k$ sourced boundaries as a sum of
falling factorials $(\alpha)_{|\pi|}$ with $|\pi|\leq k$, multiplied by
matter amplitudes independent of $\alpha$. Its absolute value is
therefore bounded by $C(1+\alpha)^m$, for a finite constant $C$
depending only on $\ell$ and $p_\text{non-SOS}$.
For integers $\alpha>M$, we also have
\begin{equation}
 0\leq s_M(\alpha)
 =\alpha^2\binom{\alpha-2}{M-1}^{\!2}
 \leq\alpha^2 4^{\alpha-2}.
 \label{eq:app-persistence-filter-bound}
\end{equation}
The remaining contribution in \eqref{eq:app-persistence-tail} thus
obeys
\begin{align}
 &\left|
 L_G\!\left[s_M(Z(1))p_\text{non-SOS}(\mathbf Z)\right]
 -\lambda e^{-\lambda}\ell(p_\text{non-SOS}(\mathbf x))
 \right|
 \nonumber\\
 &\qquad\leq\frac{Ce^{-\lambda}}{16}
 \sum_{\alpha>M}\frac{(4\lambda)^\alpha}{\alpha!}
       \alpha^2(1+\alpha)^m
 \underset{M\to\infty}{\longrightarrow}0.
 \label{eq:app-persistence-limit}
\end{align}
The factorial ensures convergence for every fixed
$0<\lambda<\infty$. Since
$\lambda e^{-\lambda}\ell(p_\text{non-SOS})<0$, then $L_G\!\left[s_M(Z(1))p_\text{non-SOS}(\mathbf Z)\right]<0$, showing the violation of statistical positivity
follows for sufficiently large but finite $M$. The non-SOS negative amplitude  involves at most $2M+m$ boundaries. No gravitational amplitudes have been
modified, so all closed-universe Gram matrices remain positive
semidefinite by \eqref{eq:app-master}. The argument uses matter moments
only through degree $m$ and imposes no growth condition on the higher
moments.

In particular, the matter amplitudes in \eqref{eq:app-seed} satisfy
$\ell(p_{\rm CL})=6-\eta/4<0$ for any fixed $\eta>24$. Hence, at every
finite positive $\lambda$, sufficiently many additional sourceless
boundaries give
\begin{equation}
 L_G\!\left[
 s_M(Z[1])p_{\rm CL}\bigl(Z[x_1],\ldots,Z[x_4]\bigr)
 \right]<0.
 \label{eq:app-persistence-choi-lam}
\end{equation}
Thus the matter interaction need not be increased with $\lambda$ to
violate statistical positivity; instead, the number of additional
sourceless boundaries may depend on $\lambda$. This establishes
persistence of the violation without requiring the original
four-boundary Choi--Lam combination to remain negative.\\

\section{A simple model where strong commutativity fails}

\label{sec:what-fails-in-the-MM-construction}

The purpose of this appendix is to explain how, in an elementary quantum mechanical example, non-SOS positivity can fail despite using a set of operators that commute. Said more carefully, we will define a matrix $M_0$ of expectation values of two operators $p_x, p_y$ such that $[p_x,p_y] = 0$. If $p_x$ and $p_y$ were ordinary real numbers, this matrix $M_0(p_x, p_y)$ would be a positive semi-definite matrix for any values of $p_x$ and $p_y$. Nevertheless, we will explicitly find a quantum state where this matrix fails to be positive semi-definite.

On the double cover of the punctured plane, take
\begin{align} \label{doubleCover}
  0\leq\theta<4\pi,
  \qquad
  \psi(r,\theta+2\pi)=-\psi(r,\theta),
  \qquad
  p_x=-i\partial_x,\quad p_y=-i\partial_y.
\end{align}
Although $p_xp_y\psi=p_yp_x\psi$ away from $r=0$~\cite{schmüdgen2026domaincharacterizationsstrongcommutativitybreak}, one cannot simultaneously diagonalize both operators. To see the obstruction, note that %
\es{}{
e^{ixp_x}e^{iyp_y}\neq e^{iyp_y}e^{ixp_x},
} 
because a closed loop around $r=0$ produces a phase. 
We can equivalently think of the wavefunction as living on a two-sheeted Riemann surface, with a branch cut from $r=0$ to $r=\infty$. As a simple diagnostic, consider a state which satisfies \eqref{doubleCover} and contains a localized wavepacket, say at $(x,y) = (0.5,0.5)$ on the first sheet (and therefore has a localized wavepacket at $(x,y)=(0.5,0.5)$ on the second sheet with the opposite sign.) Then consider the action of $e^{\i  p_y}  e^{ i p_x} e^{-i p_y}e^{-i p_x}$, which swaps the wavepacket from the first and second sheets. This gives %
\begin{align}
  \bra{\psi} e^{\i  p_y}  e^{ i p_x} e^{-i p_y}e^{-i p_x} \ket{\psi} = -1.
\end{align}
However, if we could simultaneously diagonalize all of the above operators, we could only get $+1$.

\subsubsection{ A positive-semidefinite matrix that is non-SOS} 
The above example already demonstrates a subtle failure of positivity for exponential operators. However, the gravitational path integral naturally computes moments of $\widehat{Z}_i$ operators. Therefore, we search for a quantum system where non-SOS positivity fails due to a failure of strong commutativity that only involves polynomials of the commuting operators.

Let $\mathfrak{M}$ denote the two-sheeted Riemann surface described above. We were not able to find an example of a state in $\mathcal{L}^2(\mathfrak{M})$ that violates non-SOS positivity; as argued below, the simplest example seems to involve a particle with an internal degree of freedom described by the Hilbert space $\mathcal{L}^2(\mathfrak{M}) \otimes \mathbb{C}^3$.

The example is based on the following observation.
We claim that for $k_x, k_y \in \mathbb{R}$, the following matrix is positive semi-definite, but cannot be written as a matrix sum-of-squares: $M_0 \ne \sum_I F_I^2$,
\begin{align}
 M_0=
 \begin{pmatrix}
 k_x^2+1&-k_x k_y&-k_x\\
 -k_x k_y&k_x^2+k_y^2&-k_y\\
 -k_x&-k_y&1+k_y^2
 \end{pmatrix}.
\end{align}
 To check that $M_0$ is positive semi-definite, note that
 \begin{enumerate}
     \item The diagonal entries of $M_0$ are manifestly positive.
     \item The determinant of the upper left 2$\times$2 matrix is $k_x^4 + k_x^2 + k_y^2 \ge 0$. 
     \item The full determinant 
\begin{align}\label{detm}
 \det M_0&=k_x^4 k_y^2+k_y^4+k_x^2-3k_x^2k_y^2 \geq 0 \\ %
 k_x^4 k_y^2+k_y^4+k_x^2&\geq 3\bigl(k_x^4 k_y^2\,k_y^4\,k_x^2\bigr)^{1/3} =3k_x^2k_y^2 \label{detm2}
\end{align}
In the second line \eqref{detm2} we have used the AM-GM inequality; the result justifies $\det M_0 \ge 0$.
 \end{enumerate}  
 Hence we conclude that $M_0$ is positive semi-definite since all its principal minors are positive $M_0 \succeq 0$ as a classical matrix, e.g., if we treat $k_x,k_y$ as valued in $\mathbb{R}$.
However, $M_0$ is not a matrix sum of squares.  Indeed,
\begin{align}
 M_0=F^*F
 \quad\Longrightarrow\quad
 \det M_0=\sum_I |\det F_I|^2,
\end{align}
whereas appending $k_z$'s to make every term in \eqref{detm} have degree six gives
\begin{align}
\det M_0 \to  k_x^4k_y^2+k_y^4 k_z^2+k_z^4k_x^2-3k_x^2 k_y^2 k_z^2.
\end{align}
This is the Motzkin polynomial, which is not a sum of squares.

 \subsubsection{An example}
 \newcommand{\sgn}{\operatorname{sgn}}
 An AI-assisted search yielded the following 3-component wavefunction that leads to non-SOS negativity. We define this wavefunction on the first sheet piecewise for each quadrant.
For $y>0$ the wavefunction is
\begin{equation}
 \Psi(x,y)=e^{-|x|-y}
 \begin{cases}
 \begin{pmatrix}
 -7-5y+14x-x^2y^2\\
 y(14+11xy)\\
 -i\bigl[15+12y-2x(2+y^3)\bigr]
 \end{pmatrix},&x>0,\\[14pt]
 \begin{pmatrix}
 -7-5y+14x+10x(x-y)+x^3\\
 14y-2x(10-9y+2y^2-5xy)\\
 iy(5y-18x-4x^2)
 \end{pmatrix},&x<0,
 \end{cases}
 \label{eq:integer-matrix-state}
\end{equation}
Then for $y<0$, we define the state
\begin{equation}
 \Psi(x,y)=
 \begin{pmatrix}
 -\psi_1(x,-y)\\
 \phantom{-}\psi_2(x,-y)\\
 -\psi_3(x,-y)
 \end{pmatrix},
 \qquad y<0,
 \label{eq:integer-matrix-state-lower}
\end{equation}
so the state is defined in quadrants. This choice is motivated by the symmetry $y \to -y$, $M_0(p_x,-p_y)=D^{-1}\,M_0(p_x,p_y)\,D$ where
$D=\operatorname{diag}(-1,1,-1).
$ Thus $M_0$ is invariant under the combined spatial and ``internal'' transformation $(\mathsf S_y\Psi)(x,y)=D\Psi(x,-y)$.

\begin{figure}[t!]
  \centering
  \includegraphics[width=0.8\linewidth]{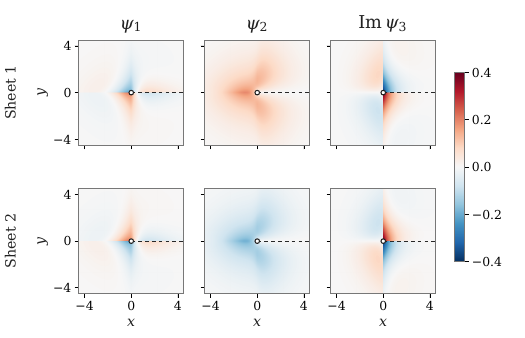}
  \caption{The wavefunction that violates non-SOS positivity. Columns depict the three components of the wavefunction; rows show the first and second sheets  with the branch cut indicated. The wavefunction is continuous across the branch cut on $\mathfrak{M}$. 
  The wavefunction changes sign between sheets.}
  \label{fig:exact-matrix-witness}
\end{figure}

The state on the second sheet is defined by $\Psi(x',y') = - \Psi(x,y)$, so we get a phase shift of $\pi$ by going around a sheet.
This is depicted in Figure \ref{fig:exact-matrix-witness}.

It is easy to verify that for this choice of $\Psi$,
\begin{align}
    \bra{\Psi} M_0 \ket\Psi = \sum_{i,j=1}^3 \int_{\mathfrak{M}} \, \psi_i^* \widehat{(M_0)_{ij}} \psi_j = -9/4
\end{align}

This wavefunction has several qualitative features. First all three components should be non-zero: if one component vanishes then $\bra{\Psi}M_0 \ket{\Psi} \ge 0$ because every $2\times2$ block of $M_0$ is a sum of squares. Each component is continuous across the branch cut. The three components also have different spatial profiles and a relative phase that allows their off-diagonal momentum couplings to outweigh the positive diagonal contributions.

\section{Non-SOS bootstrap in statistical mechanics}
\label{sec:non-SOS-bootstrap}

The overall goal of this paper is to distinguish the positivity conditions on
GPI amplitudes that follow from a Hilbert-space interpretation from the stronger conditions required for an interpretation as a genuine statistical
average.  In this appendix, we illustrate the practical strength of this
distinction in controlled statistical-mechanics examples.  We compare the
standard SOS moment bootstrap~\cite{barak2014sumofsquaresproofsquestoptimal,lasserre2018momentsoshierarchy,keren2026stationarityenoughtightnessquantum,berenstein2021bootstrappingsimpleqmsystems,Berenstein_2023,morita2026ambiguityproblembootstrapmethod}, whose feasible points can include
pseudoexpectations, with a bootstrap supplemented by positivity constraints
from explicit non-SOS polynomials.  We first study multivariate probability
distributions (for which numerical estimates for low moments are easy to obtain) and then multi-matrix integrals  (for which numerical estimates for low moments are more difficult to obtain, making the bootstrap bounds more valuable in practice).

\subsection{Bootstrapping multivariate probability distributions}
\subsubsection{A 4-variable model and the Choi-Lam Polynomial}
\newcommand{\CL}{Q_{\mathrm{CL}}}
\renewcommand{\d}{\mathrm{d}}
We first consider bootstrapping the classical statistical mechanics of a particle in a 4d potential 
\begin{align}
 Z(u,\lambda)&=\int \,d^4 x \, e^{-V(x)},\\
 V(x)&=\frac12\sum_{i=1}^4x_i^2
 +u\sum_{i=1}^4x_i^6+\lambda\CL(x),
 &\\[-2pt]
 \CL(x)&=x_1^4+x_2^2x_3^2+x_3^2x_4^2+x_4^2x_2^2
 -4x_1x_2x_3x_4.&&
\end{align}
For concreteness, for the numerics below, we choose the parameters $u=1/10$ and $\lambda = 10$. Since $Q_{\mathrm{CL}}$ is bounded below, it is clear that for $u\ge 0, \lambda \ge 0$ this potential $V$ is bounded from below and the integral converges.

\begin{figure}[h!]
  \centering
  \begin{tikzpicture}
    \begin{axis}[
      width=0.66\linewidth,
      height=3.34in,
      xmin=0,
      xmax=1.5,
      ymin=0,
      ymax=2.15,
      xlabel={$\langle x_1^2\rangle$},
      ylabel={$\langle x_1^4\rangle$},
      axis lines=box,
      tick align=inside,
      minor tick num=1,
      tick style={black,thin},
      legend style={
        at={(0.03,0.97)},
        anchor=north west,
        draw=none,
        fill=none,
        font=\small,
        row sep=-1pt
      },
      legend cell align={left},
      clip=true
    ]
      \addplot[
        draw=blue!65!black,
        fill=blue!65,
        fill opacity=0.17,
        line width=0.85pt
      ] table[x=x2,y=x4] {figures/choi_lam_level6_base.dat};
      \addlegendentry{$M_3\succeq0$ + SD}

      \addplot[
        draw=orange!82!black,
        fill=orange!85,
        fill opacity=0.28,
        line width=0.85pt
      ] table[x=x2,y=x4] {figures/choi_lam_level6_cl.dat};
      \addlegendentry{$M_3\succeq0$ + SD + $\langle Q_{\rm CL}\rangle\geq0$}

      \addplot[
        only marks,
        mark=x,
        mark size=3.3pt,
        mark options={black,line width=1.15pt}
      ] coordinates {(0.1396369284,0.0453018872)};
      \addlegendentry{numerical integral}
    \end{axis}
  \end{tikzpicture}
  \caption{The island at level 6 with and without the Choi-Lam constraint. \label{fig1} }
\end{figure}
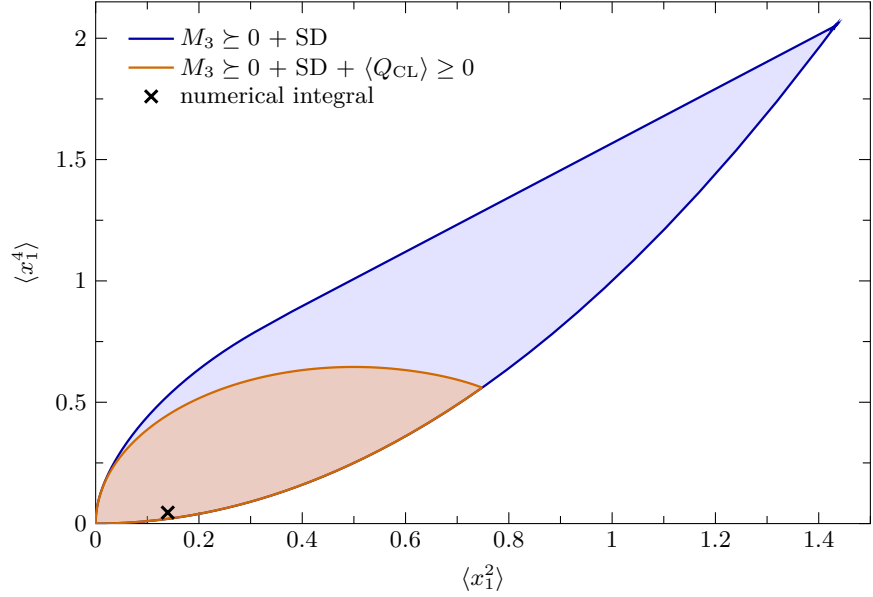

One can implement the integral bootstrap~\cite{berenstein2025goldilocksbootstrap} for this system. The equations of motion are the Schwinger-Dyson (SD) equations:
\begin{align}
0&= \frac{1}{Z}\int \d^4 x \, \partial_i \left( p(x) e^{-V}\right) = \left\langle\partial_i p-p\,\partial_iV\right\rangle\,.
\end{align}
Let us introduce the following shorthand notation for correlators:
\begin{align}
 y_{a_1 a_2 a_3 a_4} = \langle x_1^{a_1} x_2^{a_2} x_3^{a_3} x_4^{a_4} \rangle   .
\end{align}
We define the level of the bootstrap to be the maximum value of $\sum_i a_i$ for any correlator that appears in the matrices that we shall constrain by using positivity. We will consider the constraints at level 6, which means considering polynomials $P$ up to degree 3 and imposing
\begin{align}
 \langle \bar{P}(x) P(x)\rangle&\ge0\quad(\deg P\le3)\,.
\end{align}
This is equivalent to imposing the positive semi-definite constraint
\begin{align}
    M_3(y)&=\bigl(y_{\alpha+\beta}\bigr)_{|\alpha|,|\beta|\le3}\succeq0 \,, %
\end{align}
which we recall is the analog of the Hilbert space positivity condition discussed in the main text.
We also impose all Schwinger-Dyson equations involving variables up to the given level. Together, these constraints imposed up to level 6 lead to the blue allowed region for the second and fourth moments of $x_1$ in Figure \ref{fig1}.

In addition to these constraints, we can now consider imposing statistical positivity. To do this, we can impose the non-SOS constraint from the Choi-Lam polynomial:
\begin{align}\label{CLconstraint}
\langle{\CL\rangle}&=y_{4000}+y_{0220}+y_{0022}+y_{0202}-4y_{1111}\ge0.
\end{align}
This results in the orange bootstrap island in Figure \ref{fig1}. Interestingly, this constraint reduces the size of the island by a significant fraction, signaling the power of non-SOS positivity, which can thus provide significant improvements to bootstrap bounds. To further exemplify this, we discuss a statistical ensemble with fewer variables next.

\subsubsection{A 2-variable model and the Motzkin polynomial}
We can also consider a model with just two coordinates.
\begin{align}
 Z_M&=\int \d x\,\d y\, e^{-V_M(x,y)}
 &V_M(x,y)&=\frac12(x^2+y^2)+M(x,y),\label{eq:motzkin-model}%
\end{align}
Here $M$ is the Motzkin polynomial, a positive non-SOS polynomial
\begin{align}
 M(x,y)&=x^4y^2+x^2y^4-3x^2y^2+1,
 &M&\notin\Sigma[x,y]_6,\label{eq:motzkin}\\[-2pt]
 x^4y^2+x^2y^4+1&\ge3(x^6y^6)^{1/3}=3x^2y^2,
 &M(x,y)&\ge0.\label{eq:motzkin-positive}
\end{align}
At level $6$, symmetry and the Schwinger--Dyson equation give
\begin{align}
 0&=\left\langle\partial_xx-x\,\partial_xV_M\right\rangle
   =1-\langle x^2\rangle
    -6\bigl(\langle x^4y^2\rangle-\langle x^2y^2\rangle\bigr),
 &\langle x^4y^2\rangle
 &=\langle x^2y^2\rangle+\frac{1-\langle x^2\rangle}{6},
 \label{eq:motzkin-sd}
 \end{align}
Positivity gives
\begin{align}
 M_3(y)&\succeq0\quad\Longrightarrow\quad
 \begin{pmatrix}
 \langle x^2\rangle&\langle x^2y^2\rangle\\
 \langle x^2y^2\rangle&\langle x^4y^2\rangle
 \end{pmatrix}\succeq0,&&\label{eq:motzkin-minor-matrix}\\[-2pt]
 \langle x^2y^2\rangle^2
 &\le \langle x^2\rangle
 \left(\langle x^2y^2\rangle+\frac{1-\langle x^2\rangle}{6}\right).
 &&\label{eq:motzkin-minor}
\end{align}
We can analytically derive the allowed region subject to these constraints.
The blue region in Figure~\ref{fig:motzkin} is allowed by these SOS and
Schwinger--Dyson constraints, which gives a non-compact peninsula.
The peninsula is bounded below by
$\langle x^2y^2\rangle\ge\frac12\left(\langle x^2\rangle-\sqrt{\frac{\langle x^2\rangle(\langle x^2\rangle+2)}{3}}\right)$
for $\langle x^2\rangle>1$, and above by
$\langle x^2y^2\rangle\le\frac12\left(\langle x^2\rangle+\sqrt{\frac{\langle x^2\rangle(\langle x^2\rangle+2)}{3}}\right)$
for $\langle x^2\rangle>0$.

Motzkin positivity adds the new constraint
\begin{equation}
 \langle M\rangle
 =\frac{4-\langle x^2\rangle}{3}-\langle x^2y^2\rangle\ge0.
 \label{eq:motzkin-cut}
\end{equation}
Combined with \eqref{eq:motzkin-minor}, this cuts off the unbounded tail
and gives the orange region in Figure~\ref{fig:motzkin}, a bounded island with
\begin{equation}
 0\le\langle x^2\rangle\le\frac{32}{11},
 \qquad 0\le\langle x^2y^2\rangle\le1.
 \label{eq:motzkin-extrema}
\end{equation}

One can also perform the integrals numerically using the quadrature method, which gives
\begin{align}
 \langle x^2\rangle&=0.7174604370,
 &\langle x^2y^2\rangle&=0.3632609992,\label{eq:motzkin-numerics}\\[-2pt]
 \langle M\rangle&=0.7309188551,
 \label{eq:motzkin-checks}
\end{align}

\subsubsection{More positive, non-SOS polynomials}
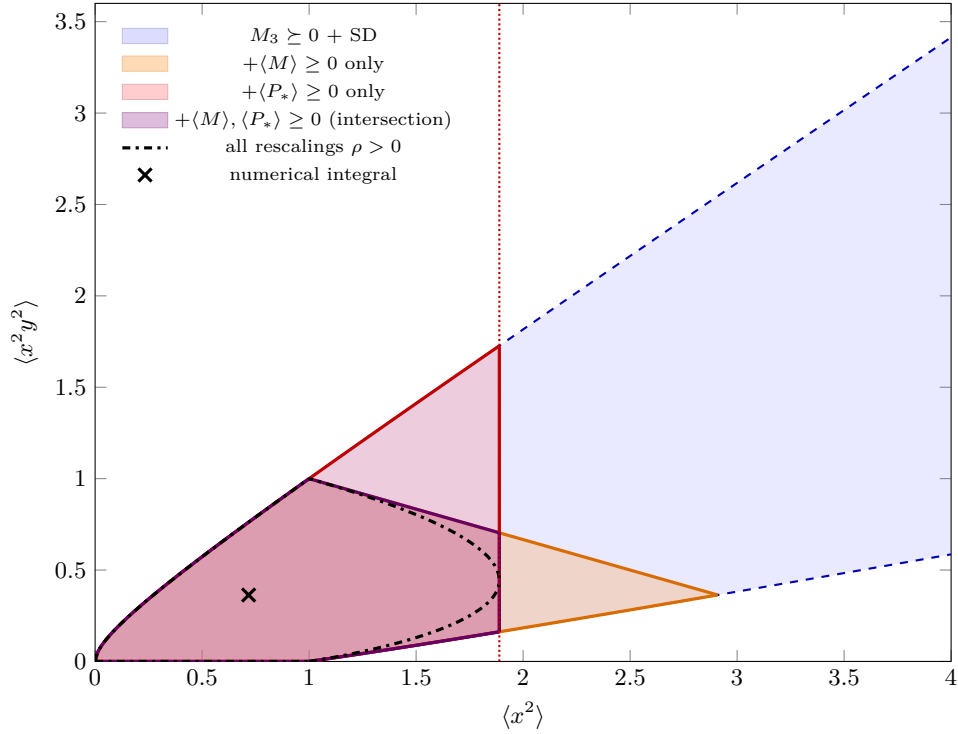
\begin{figure}[h]
    \centering
        \begin{tikzpicture}
        \begin{axis}[
          width=0.72\linewidth,height=4.05in,
          xmin=0,xmax=4,ymin=0,ymax=3.6,
          xlabel={$\langle x^2\rangle$},ylabel={$\langle x^2y^2\rangle$},
          axis lines=box,tick align=inside,
          legend style={draw=none,fill=none,font=\scriptsize,at={(0.02,0.98)},anchor=north west},
          samples=180,domain=0:4,
        ]
        \addplot[name path=baseup,draw=blue!70!black,dashed,thick,forget plot]
          {(x+sqrt(x*(x+2)/3))/2};
        \addplot[name path=baselo,draw=blue!70!black,dashed,thick,forget plot]
          {max(0,(x-sqrt(x*(x+2)/3))/2)};
        \addplot[blue!55,opacity=0.16,forget plot] fill between[of=baseup and baselo];
        \addlegendimage{area legend,fill=blue!55,draw=blue!70!black,opacity=0.25}
        \addlegendentry{$M_3\succeq0$ + SD}
        \addplot[name path=motup,draw=orange!85!black,very thick,domain=0:2.9090909,forget plot]
          {min((x+sqrt(x*(x+2)/3))/2,(4-x)/3)};
        \addplot[name path=motlo,draw=orange!85!black,very thick,domain=0:2.9090909,forget plot]
          {max(0,(x-sqrt(x*(x+2)/3))/2)};
        \addplot[orange!75,opacity=0.28,forget plot] fill between[of=motup and motlo];
        \addlegendimage{area legend,fill=orange!75,draw=orange!85!black,opacity=0.38}
        \addlegendentry{$+\langle M\rangle\ge0$ only}
        \addplot[name path=starup,draw=red!75!black,very thick,domain=0:1.8888889,forget plot]
          {(x+sqrt(x*(x+2)/3))/2};
        \addplot[name path=starlo,draw=red!75!black,very thick,domain=0:1.8888889,forget plot]
          {max(0,(x-sqrt(x*(x+2)/3))/2)};
        \addplot[red!65,opacity=0.20,forget plot] fill between[of=starup and starlo];
        \addplot[red!75!black,very thick,forget plot]
          coordinates {(1.8888889,0.1620507) (1.8888889,1.7268382)};
        \addlegendimage{area legend,fill=red!65,draw=red!75!black,opacity=0.30}
        \addlegendentry{$+\langle P_*\rangle\ge0$ only}
        \addplot[name path=smallup,draw=violet!80!black,very thick,domain=0:1.8888889,forget plot]
          {min((x+sqrt(x*(x+2)/3))/2,(4-x)/3)};
        \addplot[name path=smalllo,draw=violet!80!black,very thick,domain=0:1.8888889,forget plot]
          {max(0,(x-sqrt(x*(x+2)/3))/2)};
        \addplot[violet!65,opacity=0.30,forget plot] fill between[of=smallup and smalllo];
        \addplot[violet!80!black,very thick,forget plot]
          coordinates {(1.8888889,0.1621) (1.8888889,0.7037037)};
        \addlegendimage{area legend,fill=violet!65,draw=violet!80!black,opacity=0.40}
        \addlegendentry{$+\langle M\rangle,\langle P_*\rangle\ge0$ (intersection)}
        \addplot[black,very thick,dashdotted,domain=0:1,samples=180,forget plot]
          ({1+6*x-6*x^(3/2)},x);
        \addplot[black,very thick,dashdotted,domain=0:1,samples=180,forget plot]
          ({(6*x+1-sqrt(12*x^2+12*x+1))/2},x);
        \addplot[black,very thick,dashdotted,forget plot]
          coordinates {(0,0) (1,0)};
        \addlegendimage{black,very thick,dashdotted}
        \addlegendentry{all rescalings $\rho>0$}
        \addplot[only marks,mark=x,mark size=3.5pt,very thick,black,forget plot]
          coordinates {(0.7174604370,0.3632609992)};
        \addlegendimage{only marks,mark=x,mark size=3.5pt,very thick,black}
        \addlegendentry{numerical integral}
        \draw[red!75!black,densely dotted,thick]
          (axis cs:1.8888889,0)--(axis cs:1.8888889,3.6);
        \end{axis}
        \end{tikzpicture}
\caption{Level-6 boundaries for the two-variable model. The blue ``peninsula'' region comes from imposing positivity of SOS polynomials. Imposing positivity of non-SOS polynomials gives instead a compact island. The red and orange regions come from imposing positivity of two different non-SOS polynomials. The dashed black line comes from imposing positivity of a continuous family of non-SOS polynomials, see equation \eqref{eq:weighted-motzkin}. }
\label{fig:motzkin}
\end{figure}

It might seem rather contrived to consider potentials $V$ that involve the very same polynomial which leads to the non-SOS positivity constraint. Here we show that this is not crucial. Indeed for any $\rho >0$, we can obtain a new polynomial which is also non-SOS by a rescaling:
\begin{align}
 P_\rho(x,y)
 &=\rho^3M\!\left(\frac{x}{\sqrt\rho},\frac{y}{\sqrt\rho}\right)
 =x^4y^2+x^2y^4-3\rho x^2y^2+\rho^3\ge0.\label{eq:weighted-motzkin}
\end{align}
After imposing the Schwinger-Dyson equations, this is reduced to
\begin{align}
     \langle P_\rho\rangle
  &=(2-3\rho)\langle x^2y^2\rangle
  +\frac{1-\langle x^2\rangle}{3}+\rho^3.\label{eq:weighted-expectation}
\end{align}
An interesting value is $\rho=2/3$. Defining $P_*(x,y):=P_{2/3}(x,y)$, we find that its expectation is independent of $\langle x^2y^2\rangle$:
\begin{align}
 \langle P_*\rangle&=\langle P_{2/3}\rangle
 =\frac{17-9\langle x^2\rangle}{27}.\label{eq:weighted-cut}
\end{align}
Thus imposing $\langle{P_*}\rangle \ge 0$ slices the peninsula vertically, see Figure \ref{fig:motzkin}. It gives the simple constraint $\langle x^2 \rangle \le 17/9$. This leads to a different, overlapping island than the orange island obtained from imposing $\langle{M}\rangle \ge 0$.
We can find an even smaller island by imposing all possible constraints for $\rho >0$. The boundary of this region is shown by the dashed black curve. This curve is tangent to the line $\langle x^2 \rangle = 17/9$.

\newpage
\subsection{Bootstrapping multi-matrix integrals}
\label{sec:matrix-non-SOS-bootstrap}

Next, we discuss how statistical positivity can strengthen the matrix-integral bootstrap~\cite{Lin_2020, Kazakov_2022, Anderson:2016rcw}.  The standard moment matrix only tests positivity on sums of Hermitian squares (Hermitian-SOS).  A positive matrix measure obeys further constraints: every matrix polynomial whose normalized trace is
nonnegative must have a nonnegative expectation value, including non-SOS polynomials.  We will use a matrix version of the Motzkin polynomial, which we denote as tracial Motzkin polynomial, and show that imposing non-SOS positivity produces significantly stronger bounds with respect to using only SOS positivity constraints of the same order.

Let $A$ and $B$ be $N\times N$ Hermitian matrices, denote $\operatorname{tr}\equiv\frac1N\operatorname{Tr}$ and consider the two matrix model defined by:
\begin{align}
 Z_N(\lambda)&=\int dA\,dB\,\exp\!\left[-N^2\operatorname{tr}V_\lambda(A,B)\right],\\
 V_\lambda(A,B)&=\frac12(A^2+B^2)+\lambda M_{\rm Motz}(A,B),\label{eq:matrix-model}\\
 M_{\rm Motz}(A,B) &=AB^4A+BA^4B-3AB^2A+\mathbf 1 .
 \label{eq:matrix-motzkin}
\end{align}
The tracial Motzkin polynomial $M_{\rm Motz}$ is not positive semidefinite as a matrix but its trace is nevertheless nonnegative for every $N$.  Indeed, set $X=A^2$ and $Y=B^2$, and
let $x_i\geq0$ and $y_j\geq 0$ be their eigenvalues; we have %
\begin{align}
 \operatorname{tr} M_{\rm Motz}(A,B)
 =\frac1N\sum_{i,j}|\langle i|j\rangle|^2
 \left(x_i y_j^2+x_i^2y_j-3x_i y_j+1\right)\geq0 .
 \label{eq:matrix-motzkin-trace-positive}
\end{align}
The expression in parentheses is the Motzkin polynomial \eqref{eq:motzkin-polynomial} evaluated at $(\sqrt{x_i},\sqrt{y_j})$ which is positive.  Thus
$M_{\rm Motz}$ is \emph{trace positive}. However, it cannot be cyclically equivalent\footnote{Meaning equivalent up to terms that vanish upon taking a trace.} to a Hermitian-SOS: if it were, then choosing matrices $A=a \mathbf 1,B=b\mathbf 1$ would result in an SOS representation of the scalar Motzkin polynomial, which is impossible.  \\
A well-known theorem for matrix polynomials states that if a matrix polynomial is PSD then it is equal to a Hermitian-SOS~\cite{2002helton}. This is the reason one must work with tracial positivity instead of matrix PSD to distinguish SOS and statistical positivity. Useful references on tracial positivity, optimization, and the moment problem for non-commuting random variables include~\cite{Klep_2008,burgdorf2010truncatedtracialmomentproblem,Klep_2021,Pironio_2010,Huber_2024} and the reviews~\cite{helton2006positivepolynomialsscalarmatrix,BurgdorfKlepPovh2016}.

The Gaussian term in~\eqref{eq:matrix-model} is essential because the pure Motzkin potential is flat, for example, along $B=0$.  By
\eqref{eq:matrix-motzkin-trace-positive},
\begin{align}
 \operatorname{tr} V_\lambda(A,B)
 \geq\frac12\operatorname{tr}(A^2+B^2),
 \label{eq:matrix-stability}
\end{align}
so the finite-$N$ integral converges.  We take $\lambda=1/2$ in all the
numerical results below and restrict to the planar phase-preserving
$A\mapsto-A$, $B\mapsto-B$, and $A\leftrightarrow B$.  These transformations
generate the $D_4$ symmetry used to reduce the bootstrap equations.

For a word $w(A,B)$ in the $A,B$ matrices, we use $\langle\operatorname{tr} w(A,B)\rangle$.
If the basis contains words $u,v$ of length at most $ d$, its Gram matrix contains moments through length $L=2d$.  We call $L$ the bootstrap level and impose
\begin{align}
 \Gamma^{(d)}_{u,v}&=\langle\operatorname{tr} u^\ast v\rangle\succeq0
 ,\qquad |u|,|v|\leq d.
 \label{eq:matrix-moment-matrix}
\end{align}
This condition is precisely $\langle\operatorname{tr} q^\ast q\rangle\geq0$ for all polynomials of degree at most $d$; it plays the same role as the scalar moment matrix in the previous section. We then impose the planar loop equations (e.g. we work in the 't Hooft limit $N\to \infty$ with $\lambda$ held fixed, and impose large $N$ factorization \cite{Lin_2020,Kazakov_2022}):
\es{}{
\left\langle\operatorname{tr}\!\left(w\,\partial_A^{\rm cyc}V_\lambda\right)\right\rangle
 &=\sum_{w=uAv}\langle\operatorname{tr} u\rangle\langle\operatorname{tr} v\rangle,
 \label{eq:matrix-loop-equation}
}
together with the analogous equations obtained by taking derivatives with respect to $B$ and $D_4$ symmetry. We solve this system at levels $L=6,8,10$.

\begin{figure}[htbp]
\centering
\includegraphics[width=\textwidth]{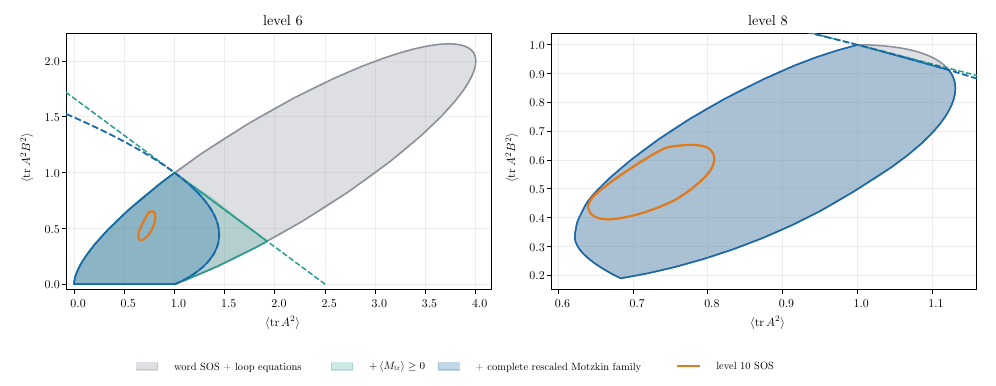}
\caption{Matrix bootstrap for the Motzkin-potential model at $\lambda=1/2$. At $L=6$ (left), the family of rescaled polynomials (blue area) removes $74.3\%$ of the SOS constraint's gray area.  At $L=8$
(right), it removes only $1.13\%$ with respect to the SOS constraints; the orange curve shows the level $L=10$ SOS 
bound for control. }
\label{fig:matrix-bootstrap-regions}
\end{figure}

The observables we study are
\begin{align}
 &\langle\operatorname{tr} A^2\rangle,\qquad
 \langle\operatorname{tr} A^2B^2\rangle,\qquad
 \langle\operatorname{tr} A^2B^4\rangle=\langle\operatorname{tr} A^4B^2\rangle.
 \label{eq:matrix-observables}
\end{align}
These are, respectively, the second moment of either matrix, the lowest mixed
single-trace correlator, and the degree-six moment entering the
Motzkin inequality.  At $L=6$ and $\lambda=1/2$, the only independent loop
equation and the unscaled Motzkin constraint give
\es{}{
1&=\langle\operatorname{tr} A^2\rangle+3(\langle\operatorname{tr} A^2B^4\rangle-\langle\operatorname{tr} A^2B^2\rangle),\label{eq:matrix-basic-projected-cut}\\
\langle\operatorname{tr} M_{\rm Motz}\rangle=&1+2\langle\operatorname{tr} A^2B^4\rangle-3\langle\operatorname{tr} A^2B^2\rangle\geq0.
}
The single condition $\langle\operatorname{tr} M_{\rm Motz}\rangle\geq0$ at level $L=6$ reduces the area of the allowed region from $3.530$ to $1.088$, or by $69.2\%$.
We can obtain a stronger bound by rescaling the two matrices.  Applying
\eqref{eq:matrix-motzkin-trace-positive} to $aA,bB$ gives
\begin{align}
 1+a^2b^4\langle\operatorname{tr} A^2B^4\rangle+a^4b^2\langle\operatorname{tr} A^2B^4\rangle-3a^2b^2\langle\operatorname{tr} A^2B^2\rangle\geq0,
 \qquad a,b>0 .
 \label{eq:matrix-rescaled-motzkin}
\end{align}
At fixed $a^2b^2$, the positive degree-six terms are minimized at $a=b=s$.
Writing $t=s^2$, the minimum of $1+2\langle\operatorname{tr} A^2B^4\rangle t^3-3\langle\operatorname{tr} A^2B^2\rangle t^2$ occurs at $t=\langle\operatorname{tr} A^2B^2\rangle/\langle\operatorname{tr} A^2B^4\rangle$ and is
$1-\langle\operatorname{tr} A^2B^2\rangle^3/\langle\operatorname{tr} A^2B^4\rangle^2$.  Hence the complete two-parameter rescaling family in the
$D_4$-symmetric sector is equivalent to
\begin{align}
 \langle\operatorname{tr} A^2B^2\rangle^3&\leq \langle\operatorname{tr} A^2B^4\rangle^2,
 &\langle\operatorname{tr} A^2B^2\rangle,\langle\operatorname{tr} A^2B^4\rangle&\geq0,\label{eq:matrix-power-cone}\\[-2pt]
 \langle\operatorname{tr} A^2B^2\rangle^3&\leq\left(\langle\operatorname{tr} A^2B^2\rangle+\frac{1-\langle\operatorname{tr} A^2\rangle}{3}\right)^2
 \qquad (L=6).
 \label{eq:matrix-rescaled-projected-cut}
\end{align}
The results are shown in Figure ~\ref{fig:matrix-bootstrap-regions} in the left plot. 
At $L=6$, the full rescaled family lowers the bounded area from $3.530$ to $0.906$, a reduction of $74.3\%$.
The corresponding upper bounds improve from
\begin{align}
 \langle\operatorname{tr} A^2\rangle_{\text{upper}}:4&\longrightarrow\frac{13}{9},
 &\langle\operatorname{tr} A^2B^2\rangle_{\text{upper}}:2.155&\longrightarrow1 .
 \label{eq:matrix-coordinate-bounds}
\end{align}
The first bound follows analytically from
\begin{align}
 \langle\operatorname{tr} A^2\rangle&=1+3(\langle\operatorname{tr} A^2B^2\rangle-\langle\operatorname{tr} A^2B^4\rangle)\nonumber\\
 &\leq1+3(\langle\operatorname{tr} A^2B^2\rangle-\langle\operatorname{tr} A^2B^2\rangle^{3/2})\leq\frac{13}{9},
 \label{eq:matrix-Smax-proof}
\end{align}
with equality at $\langle\operatorname{tr} A^2B^2\rangle=4/9$, $\langle\operatorname{tr} A^2B^4\rangle=8/27$.  For the second, moment-matrix
positivity gives $\langle\operatorname{tr} A^2B^2\rangle^2\leq \langle\operatorname{tr} A^2\rangle\,\langle\operatorname{tr} A^2B^4\rangle$.  
Figure~\ref{fig:matrix-bootstrap-regions}
shows the nested regions; clearly, the non-SOS constraint leads to a significant improvement over SOS constraints at fixed level.

\begin{figure}[htbp]
\centering
\includegraphics[width=0.88\textwidth]{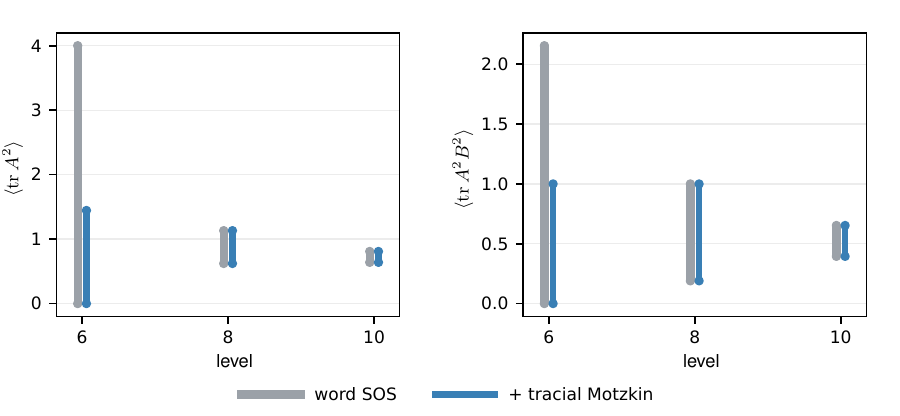}
\caption{Strength of the non-SOS constraints with increasing bootstrap level.}
\label{fig:matrix-bootstrap-hierarchy}
\end{figure}

\textbf{Increasing the bootstrap level.}
A particularly direct diagnostic of the efficacy of the non-SOS constraint is to minimize $\langle\operatorname{tr} M_{\rm Motz}\rangle$ in the ordinary
SOS bootstrap, without imposing trace positivity.  A negative value corresponds to a
pseudo-expectation which obeys all truncated Gram and loop equations but
cannot arise from an actual pair of Hermitian matrices.  We find
\begin{align}
\begin{array}{c|ccc}
 L&6&8&10\\ \hline
 \min_{\mathrm{ordinary}}\langle\operatorname{tr} M_{\rm Motz}\rangle
 &-3.070361&-0.029567&0.489413
\end{array}.
\label{eq:matrix-motzkin-minima}
\end{align}
The violation is large at $L=6$ where the non-SOS positivity constraint is strongest, smaller at $L=8$, and absent at
$L=10$.  Correspondingly, the rescaled Motzkin family removes $74.3\%$ and
$1.13\%$ of the tested reconstructed regions at the first two levels.  At
$L=10$ no improvement is obtained.
Figure~\ref{fig:matrix-bootstrap-hierarchy} summarizes the bounds on the second and mixed fourth moments.

The trace inequality~\eqref{eq:matrix-motzkin-trace-positive} is exact at
every finite $N$, whereas the numerical bootstrap assumes planar factorization with unbroken $D_4$ symmetry.  Note that at $L=10$, the loop equations contain the term $z= \langle\operatorname{tr} A^2\rangle^2$ which can be dealt with via non-linear relaxation \cite{Kazakov_2022} by introducing a new variable $z\geq \langle\operatorname{tr} A^2\rangle^2$.  
We stress that trace positivity of $M_{\rm Motz}$ does not imply the PSD 
 property $M_{\rm Motz}(A,B)\succeq0$, and therefore it would be incorrect to impose 
stronger constraints such as $\langle\operatorname{tr} R^\ast M_{\rm Motz}R\rangle\geq0$.

The bootstrap bounds can presumably be improved by including additional, higher-degree families of non-SOS trace-positive matrix polynomials~\cite{Klep_2022, Klep_2021,BurgdorfKlepPovh2016}.  The main
lesson of this simple example is that, at a fixed bootstrap cutoff, statistical non-SOS positivity can sharply strengthen the matrix bootstrap bounds with respect to only imposing SOS positivity.

\bibliographystyle{apsrev4-2-titles}
\bibliography{biblio_shortpositivity}

\end{document}